\documentclass{chet}

\usepackage{graphicx}
\usepackage[crop=pdfcrop]{pstool}
\usepackage{float}
\usepackage{mathrsfs}
\usepackage{dsfont}
\usepackage{amsfonts}

\newcommand{\vev}[1]{\langle {#1}\rangle}

\newcommand{\pd}{\partial}
\newcommand{\<}{\langle}
\renewcommand{\>}{\rangle}

\newcommand{\CO}{\mathcal{O}}
\newcommand{\CJ}{\mathcal{J}}

\newcommand{\half}{\frac{1}{2}}

\newcommand{\osbx}[2]{x_{#1\hspace{1pt}}{\!}^{#2}}
\newcommand{\IZ}{\mathbb{Z}}
\DeclareMathOperator{\Ima}{Im}
\DeclareMathOperator{\Rea}{Re}
\DeclareMathOperator{\Tr}{Tr}
\DeclareMathOperator{\Str}{Str}
\DeclareMathOperator{\Disc}{Disc}
\DeclareMathOperator{\Li}{Li}

\DeclareMathOperator{\FT}{F\!.\!T\!.}

\preprint{UCSD-PTH-11-09}
\title{Superconformally Covariant OPE and General Gauge Mediation}
\author{Jean-Fran\c{c}ois Fortin, Kenneth Intriligator and Andreas Stergiou}
\affiliation{Department of Physics, University of California, San Diego, La Jolla, CA 92093 USA}
\abstract{We consider using broken superconformal symmetry and the super operator product expansion (sOPE) to constrain and analyze hidden sector theories that couple to our gauge forces and are not necessarily weakly coupled.   Conformal and supersymmetry breaking are IR effects, associated with field or spurion expectation values, whereas the sOPE is determined in the UV and hence does not notice the breaking.  The broken superconformal symmetry relates OPE coefficients of superconformal descendant operators to those of the superconformal primaries.  We apply these ideas to the current correlators of general gauge mediation (GGM).  We also consider analyticity properties of these correlators, e.g.\ their discontinuities, and use the optical theorem to relate them to total scattering cross sections from visible to hidden sector states, e.g.\ $\sigma(\text{vis}+\text{vis}\to \text{hidden})$, analogous to $\sigma (e^+e^-\to\text{hadrons})$ in QCD.   We discuss how the current-current OPE can be truncated to the first few terms to get a good approximation to the visible sector soft masses of GGM.}

\date{September 2011}

\begin{document}

\maketitle

%%%%%%%%%%%%%%%%%%%%%%%%%%%%%%%%%%%%%%%%%%%%%%%%%%%%%%%%%%%%%%%%%%%%%%%%%%%%%%%%%%%%%%%%%%%%%%%%%%%%%%%%%%%%%%%%%%%%%%%%%%%%%%%%%%%%%%%%%%%%%%%%%%%%
%\toc
\newsec{Introduction}

Symmetries, even if they are broken, can usefully constrain theories and their dynamics.  Soft breaking can be regarded as spontaneous, even if it is actually explicit, via background or spurion expectation values.  The symmetry breaking is an IR effect, and the unbroken symmetry can still apply to constrain UV physics.   The operator product expansion (OPE) gives a particularly useful way to separate UV physics from long-distance IR physics \rcite{Wilson:1969zs}.   We will here discuss and explore applications of breaking an interesting, large symmetry group, superconformal symmetry, via the OPE.   

To set the stage,  recall how the hadronic world is probed by $e^+ e^-\to e^+ e^-$ scattering, via an intermediate photon, with the QCD contributions to the electromagnetic current two-point correlator.   Writing the current-current OPE schematically as 
\eqn{J(x)J(0)=\sum _i c_{JJ}^i {\cal O}_i,}[JJi]
the idea is that $c_{JJ}^i$ ``Wilson coefficients" are determined by UV physics, while IR physics determines the expectation values of the operators on the RHS.    Keeping only a few leading operators often suffices to obtain good qualitative insights (despite the fact that the errors in these approximations can be difficult to estimate).   There is an extensive literature on using this and related ideas to study the hadronic world, e.g.\ the classic papers of SVZ on QCD sum rules \rcite{Shifman:1978bx,Shifman:1978by}.   The UV physics can be constrained by a larger symmetry group, including broken generators.

Now consider an analogy with the above discussion where, instead of using lepton sector scattering to probe the hadronic sector, we consider scattering of our world's visible sector fields to probe a new, hidden sector, which couples to our world via gauge interactions.  The hidden sector then contributes to $SU(3)\times SU(2)\times U(1)$ current correlators, and we can try to employ the power of the OPE to separate UV vs IR physics.   The UV theory might be asymptotically free, like QCD, or an interacting, superconformal field theory (SCFT).  

Our main motivation is to apply these considerations to  general gauge mediation (GGM) \rcite{Meade:2008wd}, where indeed the visible sector soft masses are directly determined by the hidden sector's contribution to the gauge-current two-point correlators \rcite{Meade:2008wd, Buican:2008ws}:
\eqna{
M_{\rm gaugino} & =\pi i\alpha\int d^4x\,\<Q^2(J(x)J(0))\>,\\
m_{\rm sfermion}^2 & =4\pi\alpha Y\<J(x)\>+\frac{i\alpha^2c_2}{8}\int d^4x\ln(x^2M^2)\<\bar{Q}^2Q^2(J(x)J(0))\>.
}[GGMsoft]
The IR theory is neither conformal nor supersymmetric, e.g.\ because of messenger masses $M$ and mass splittings  $\sqrt{F}$.  We will explore the constraints that follow if these soft symmetry breaking effects can be regarded as spontaneous (even if they are actually explicit, via spurions), and therefore effectively restored in the UV.  In particular, we apply the UV constraints of superconformal symmetry to constrain the Wilson coefficients in the OPE \JJi\ in \GGMsoft.  The IR breaking effects then show up in the IR, via operator expectation values on the RHS of the OPE.  Even if the OPE results are only approximate, they give a foothold to consider GGM with non-weakly-coupled hidden sectors.  
  
We discussed some general aspects about the OPE of conserved currents in superconformal theories in \rcite{Fortin:2011nq}.   Leading terms at short distance include
\eqn{J_a(x) J_b(0)=\tau \frac{\delta _{ab}\mathds{1}}{ 16\pi ^4x^4}+\frac{k d_{abc}}{\tau} \frac{J_c(0)}{16\pi ^2x^2}+w\frac{\delta _{ab}K(0)}{ 4\pi ^2x^{2-\gamma _K}}+c_{ab}^i\frac{\CO_i(0)}{ x^{4-\Delta _i}}+\cdots,}[jjex]
with $a$ an adjoint index for the (say simple) group $G$; for simplicity, we will mostly take $G=U(1)$ in what follows.   The coefficient $\tau $ of the unit operator can be exactly determined from a 't Hooft anomaly  $\tau =-3\Tr RFF$\rcite{Anselmi:1997ys} using \rcite{Intriligator:2003jj} if needed, and gives  the leading coefficient of CFT ``matter"  to the $G$ gauge beta function, see e.g.\ \rcite{Barnes:2005zn}.   The  coefficient $k$ in \jjex\ of the 't Hooft anomaly  $k\sim \Tr G^3$ must vanish or be cancelled to  weakly gauge the $G$ symmetry. The operator $K$ in \jjex\ refers to an operator that classically has $\Delta =2$, e.g.\ the K\"ahler potential, but is not conserved by the interactions so it has anomalous dimension,  $\Delta _K=2+\gamma _K$.  $\CO_i(0)$ in \jjex\ is a generic, real superconformal primary, and $\cdots$ denotes other terms, including superconformal descendants.  

Superconformal symmetry together with current conservation implies that the OPE coefficients of superconformal descendants in \jjex\ are completely determined by those of the superconformal primaries \rcite{Fortin:2011nq}.\footnote{This is not as obvious as it sounds, because of the existence of the nilpotent superconformal three-point function quantities $\Theta$ and $\bar \Theta$ of \rcite{Osborn:1998qu}, see  \rcite{Fortin:2011nq} for additional discussion.}   Such relations apply in the far UV, but can be altered for example by RG running of the coefficients, because the theory is ultimately not superconformal.  Nevertheless, the UV relations of superconformal symmetry can have approximate vestiges in the IR, to be explored here. 

\begin{figure}[ht]
\centering
\includegraphics{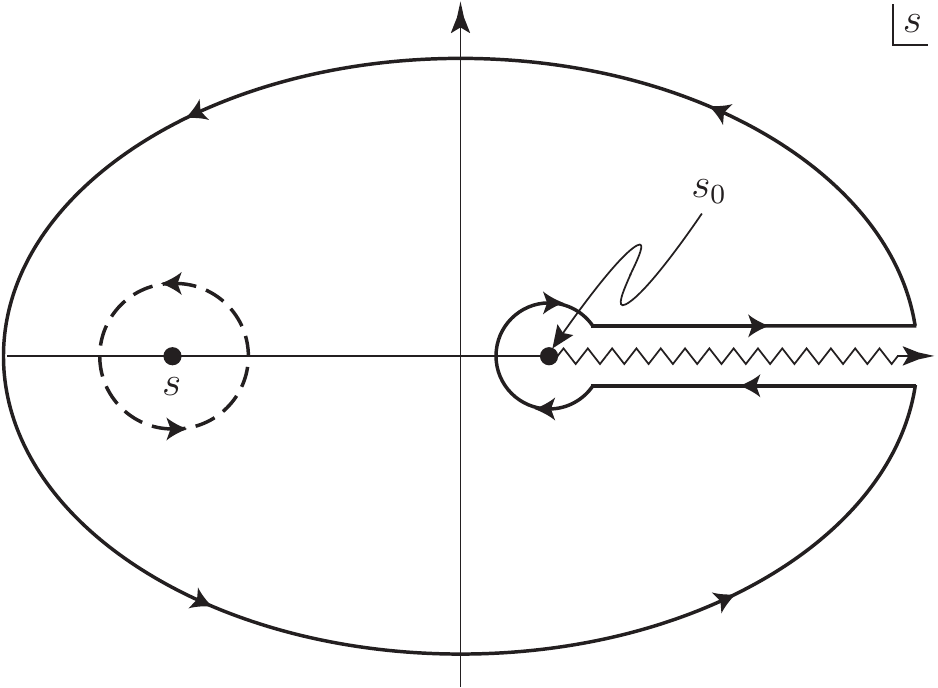}
%\pstool{./figures/cauchy}{\psfrag{o}[][]{$s_0$}
%\psfrag{s}[][]{$s$}}
\caption{The dashed contour is deformed to the solid contour, giving  \integrate. (The contribution from the circle at infinity is assumed to vanish.)}\label{cauchy}
\end{figure}

We also explore a related topic, the analyticity properties of the GGM \rcite{Meade:2008wd} current correlator functions $\widetilde C_{a=0, 1/2, 1}(p^2)$, and $\widetilde B_{1/2}(p^2)$.    These functions can have cuts when $s=-p^2$ is big enough to create on-shell states, with the cut discontinuity related by the optical theorem to total cross sections for hidden-sector state production, $\sigma _a(\text{vis}\to \text{hid}, s)$, in analogy with QCD production $\sigma (e^+e^-\to \text{hadrons})$:
\eqn{\sigma _a(\text{vis}\to \text{hid})=\frac{(4\pi \alpha ) ^2}{ s}\half \Disc \widetilde C_a(s),}[Optical]
 As in QCD applications,  we can express visible-sector observables $A(s)$ as $s$-integrals of their  discontinuity along the cut (see Fig.\ \ref{cauchy}), 
\eqn{A(s)=\frac{1}{2\pi i} \int _{s_0}^\infty ds'\, \frac{\Disc A(s')}{ s'-s}=\frac{1}{\pi}\int _{s_0}^\infty ds'\, \frac{\Ima A(s')}{ s'-s},}[integrate]
and then approximate by going to large $s^\prime$, applying the OPE, and keeping only the first few terms in the $1/s$ expansion.
We use this to show that the GGM soft masses \GGMsoft\ can be approximated in this way in terms of the lowest dimension operators appearing in \jjex\ that can have non-zero SUSY-descendant expectation values: 
\eqna{
M_{\rm gaugino} & \approx-\frac{\alpha\pi w\gamma_{Ki}}{8M^2}\<Q^2(\mathcal{O}_i(0))\>,\\
m_{\rm sfermion}^2 & \approx4\pi\alpha Y\<J(x)\>+\frac{\alpha^2c_2 w\gamma_{Ki}}{64M^2}\<\bar{Q}^2Q^2(\mathcal{O}_i(0))\>,
}[introsoft]
where $w$ is the coefficient of $K(0)$ in \jjex\ and $\gamma_{Ki}$ is the anomalous-dimension matrix which mixes $K$ with the operator $\mathcal{O}_i$.

These considerations also constrain the possibilities for GGM functions $\widetilde C_a(s)$, $\widetilde B_{1/2}(s)$.  We can use \integrate\ to relate these functions to integrals of their discontinuities (as a spectral representation), and the optical theorem \Optical\ to  relate these discontinuities to kinematic phase space factors.  For example, for producing two scalars of masses $m_1$ and $m_2$, 
\eqn{\sigma _0(s)=\frac{\lambda ^{1/2}(s, m_1, m_2)}{ 8\pi s^2}|{\cal M}|^2,}[sproduction]
where the phase-space prefactor  involves the standard (see e.g.\ \rcite{Itzykson:1980rh}) factor
\eqn{\lambda ^{1/2}(s, m_1, m_2)=2\sqrt{s}\vert\vec p\hspace{1pt}\vert=\sqrt{[s-(m_1+m_2)^2][s-{(m_1-m_2)}^2]}\,\theta (s-(m_1+m_2)^2),}[Lambdais]
where $|\vec p\hspace{1pt}|$ is the CM momentum of the produced on-shell scalars (the step function $\theta$ indicates that it is non-zero for real $s\geq (m_1+m_2)^2$).  Comparing with \Optical, 
\eqn{\Disc \widetilde C_0(s)=\frac{\lambda ^{1/2}(s, m_1, m_2)}{ 4\pi s}\left\vert\frac{{\cal M}}{ 4\pi \alpha}\right\vert^2.}[discr]

As a concrete example, consider minimal gauge mediation (MGM), with charged messenger scalars of mass $m_\pm$ and fermions of mass $m_0$.   The superpotential is  $W=hX\Phi \widetilde \Phi$, where $\Phi$, $\widetilde \Phi$ are charged messengers, with masses given by $\vev{hX}=M+\theta ^2F$, which leads to two fermions of mass $m_0=M$ and scalars of mass $m_{1,2}=m_\pm= \sqrt{|M|^2\pm |F|}$.   The functions $\widetilde C_{a}(p^2)$ of GGM have cuts where these states can go on shell, with discontinuity  related to the corresponding total cross sections as in \Optical, e.g.\ $\widetilde C_0(s)$ has a cut for $s\geq (m_++m_-)^2$, corresponding to production of the  scalars with masses $m_+$ and $m_-$, given by \discr\ with the tree-level amplitude ${\cal M}=4\pi\alpha$, so
\eqn{\half \Disc \widetilde C_0(s)=\frac{1}{ 4\pi s}\sqrt{s^2-4|X|^2s+4|F|^2},\qquad\text{for } s\geq (m_++m_-)^2,}[discXF]
 Likewise, $\widetilde C_{1/2} (s)$, $\widetilde C_1(s)$, and $\widetilde B_{1/2}(s)$ have related discontinuities.  For this example, these relations are of course readily verified from the known, explicit expressions for the GGM functions of weakly coupled theories.  But one could imagine non-weakly coupled examples, where these analyticity properties could usefully constrain the GGM functions.

In the above discussion $X$ can either be a dynamical field, the goldstino superfield, or spurion of the spurion limit.   We separate the UV description, sufficiently far above $\vev{X}$, from the IR effects of $\vev{X}$.  In the UV description, the messengers are effectively massless and interacting with $X$ with coupling $h$ (we avoid going too far in the UV, to avoid $h$'s Landau pole).   We illustrate how to reproduce e.g.\ \discXF\ from direct computations of the Wilson coefficients of the two-point OPE of the current superfield  $\CJ=\Phi ^\dagger \Phi -\widetilde \Phi ^\dagger \widetilde \Phi$, to terms on the RHS of the OPE involving  the operators $(X^\dagger X)^n$, and superconformal descendants. (Aspects of the OPE interpretation of super-propagators was explored for some interacting theories in \rcite{Leroy:1986ve}.)
As we will illustrate and verify, the superconformal symmetry implies many relations among the various terms.    In the IR, we replace $X\to \vev{X}$, and these terms then contribute to, and indeed reproduce, the  GGM \rcite{Meade:2008wd} current-current correlators. 

The paper is organized as follows: section \ref{OPE} reviews the OPE, superconformal covariance, and the results of  Ref.~\rcite{Fortin:2011nq} for current-current correlation functions in general superconformal theories.  In section \ref{GGM} we apply  these results to the general gauge mediation functions $C_a$ and $B_{1/2}$ \rcite{Meade:2008wd}, discussing how these functions can be constrained by approximate, broken, superconformal symmetry.   In section \ref{analggm} we study the analyticity properties and constraints on the GGM functions, and how the OPE can be applied to obtain approximations \introsoft\ for soft terms in theories that aren't necessarily weakly coupled.   Section \ref{MGM} illustrates and checks our various general results in the well-studied example of weakly coupled minimal gauge messenger mediation MGM.  Section \ref{CONC} summarizes and mentions possible further applications of our findings.  Appendix \ref{AppWils} illustrates explicit computations of current-current OPE Wilson coefficients in MGM.

%%%%%%%%%%%%%%%%%%%%%%%%%%%%%%%%%%%%%%%%%%%%%%%%%%%%%%%%%%%%%%%%%%%%%%%%%%%%%%%%%%%%%%%%%%%%%%%%%%%%%%%%%%%%%%%%%%%%%%%%%%%%%%%%%%%%%%%%%%%%%%%%%%%%

\newsec{The operator product expansion}[OPE]

The  OPE \rcite{Wilson:1969zs} replaces nearby operators with a sum of local operators
 \eqn{
\mathcal{O}_i(x)\mathcal{O}_j(0)={}\sum_{k}c_{ij}^k(x, P)\mathcal{O}_k(0),
}[OPEx]
where $c_{ij}^k(x, P)$ are the (position space) Wilson coefficients (with $[P_\mu, \CO]=i\partial _\mu \CO$).  In non-scale invariant theories, \OPEx\ approximately holds for small $x$, or in the light-cone limit of small $x^2$, while for CFTs \OPEx\ is exact.  In momentum space, 
\eqn{
i\int d^4x\,e^{-ip\cdot x}\mathcal{O}_i(x)\mathcal{O}_j(0)\xrightarrow[p^2\to-\infty]{} \sum_{k}\tilde{c}_{ij}^k(p)\mathcal{O}_k(0),
}[OPEp]
with the Fourier transform applied on the Wilson coefficients, while the operators $\mathcal{O}_k(0)$ remain in position space.  The coefficients $\tilde{c}_{ij}^k(p)$ can be extracted from the OPE \OPEp sandwiched between appropriate in and out external states.  

In applications of the OPE to non-scale invariant theories, e.g.\ QCD, one splits momentum integrals into UV and IR regions, above and below a renormalization scale $\mu$.  For the IR physics of the renormalized operators, in particular their expectation values, $\mu$ acts as a UV cutoff scale.  For the UV physics, namely the Wilson coefficients, $\mu$ acts as an IR cutoff scale. For a spirited discussion of the properties of the OPE, and the necessity of this splitting at a scale $\mu$, the reader is referred to \rcite{Novikov:1984rf}.    The scale $\mu$ drops out of physical quantities at the end of the day, of course. The coefficients obey an RG equation
\eqn{{\cal D}c_{ij}^k(\mu ^2 x^2)=\gamma ^k_ \ell C_{ij}^\ell -\gamma _i^\ell C_{\ell j}^k-\gamma _j^\ell C_{i\ell}^k,}[RGe]
with ${\cal D}=\mu \frac{\partial }{ \partial \mu}+\beta (g)\frac{\partial }{ \partial g}.$  Even if the theory is RG flowing, with non-zero beta functions, this is accounted for by these RG equations, making the OPE still effectively scale covariant.

\subsec{Conformal-symmetry constraints}

Exactly scale-invariant theories are generally also conformally invariant (modulo recently found counterexamples \rcite{Fortin:2011ks,Fortin:2011sz}).  We're here ultimately interested in applying the OPE also to non-scale-invariant theories, but the intuition is that the Wilson coefficients are UV-determined, so we can work near the approximately conformally invariant UV fixed point, obtain relations there, and then RG flow them down to lower scales.  The Wilson coefficients should then (approximately) respect the full conformal group, i.e.\ they should respect not only the dilation generator, $D$, but also the special conformal generator, $K_\mu$.  These generators act on primary operators  $\CO ^I$ ($I$ labels the $(j, \bar \jmath)$ spin indices) as 
\begin{gather*}
[P_\mu,\mathcal{O}^I(x)]=i\partial_\mu\mathcal{O}^I(x),\quad\quad [D,\mathcal{O}^I(x)]=-i(x\cdot\partial+\Delta_{\mathcal{O}})\mathcal{O}^I(x),\displaybreak[0]\\
[K_\mu,\mathcal{O}^I(x)]=i(x^2\partial_\mu-2x_\mu\,x\cdot\partial-2\Delta_{\mathcal{O}}x_\mu)\mathcal{O}^I(x) +2x^\nu(s_{\mu\nu})_{\!\phantom{I}J}^I\mathcal{O}^J(x),
\end{gather*}
where $(s_{\mu\nu})_{\!\phantom{I}J}^I$ is the operator's Lorentz spin representation,  and $\Delta_{\mathcal{O}}$ is its scaling dimension.  

Conformal symmetry implies that the  OPE of conformal descendants are fully determined by those of the conformal primaries   \rcite{Ferrara:1973yt}. For example, for the OPE of two scalar operators, \eqn{\CO_i(x_1)\CO _j(x_2)=\sum 
_{\substack{\rm{primary}\\ \CO ^{(\ell)}_k}} c_{ij}^k\frac{1}{ r_{12}^{\half (\Delta _i+\Delta _j-\Delta _k)}}F_{\Delta _i\Delta _j}^{\Delta _k}(x_{12}, P)_{\mu _1\dots \mu _\ell}\CO _k^{(\mu _1\dots \mu _\ell)}(x_2),}[OPEspin]
where $x_{ij}\equiv x_i-x_j$ and $r_{ij}\equiv x_{ij}^2$ and the sum is over  integer spin-$\ell$ primary operators  $\CO _k^{(\mu _1\dots \mu _\ell)}$ (with symmetrized indices) on the RHS.  The functions $F_{\Delta _i\Delta _j}^{\Delta _k}(x_{12}, P)_{\mu _1\dots \mu _\ell}$, which give the coefficients of the descendants, are fixed by  conformal covariance.   Equivalently, conformal symmetry completely fixes the form of the two-point and three-point functions up to an overall coefficient.  For example, the three-point functions related to \OPEspin\ are 
\eqn{\vev{\CO _i(x_1)\CO _j (x_2)\CO _k^{(\mu _1\dots \mu _\ell)}(x_3)}=\frac{c_{ijk}}{ r_{12}^{\half (\Delta _i+\Delta _j-\Delta _k+\ell)}r_{13}^{\half (\Delta _k+\Delta _{ij}-\ell)}r_{23}^{\half (\Delta _k-\Delta _{ij}-\ell)}}Z^{(\mu _1}Z^{\mu _2}\cdots Z^{\mu _\ell)},}[OOOspin]
where  $\Delta _{ij}\equiv \Delta _i-\Delta _j$, and 
\eqn{Z^\mu \equiv \frac{x_{23}^\mu }{ r_{23}}-\frac{x_{13}^\mu }{ r_{13}}, \qquad Z^2=\frac{r_{12}}{ r_{13}r_{23}}.}[Zis]

We'll sometimes be interested in Fourier transforming the OPEs, as in \OPEp.   For example, in \OPEspin, taking $x_2=0$ and Fourier transforming in $x_1\equiv x$,
\eqn{i\int d^4 x\, e^{-ip\cdot x}\CO_i(x)\CO _j(0)\supset c_{ij}^kF_{\Delta _i\Delta _j}^{\Delta _k}(-i\partial _p, P)_{\mu _1\dots \mu _\ell} \FT\left(\frac{1}{ (x^2)^{\half (\Delta _i+\Delta _j-\Delta _k)}}\right)\CO _k^{(\mu _1\dots \mu _\ell)}(0),}[FTOPEspin]
 The Fourier integral is generally singular but can be defined by analytic continuation, with
 \eqn{\FT \left(\frac{1}{ (x^2)^d}\right)\equiv i\int d^4 x\, e^{-ip\cdot x}\frac{1}{ (x^2)^d}=(2\pi )^2\frac{\Gamma (2-d)}{ 4^{d-1}\Gamma (d)}(p^2)^{d-2}.}[FourTr]
Logarithms of $p^2$ can arise if the dimension $d$ is an integer $n$, or nearby, $d=n+\epsilon$, with $\epsilon\ll 1$.  The $1/\epsilon$ terms are local contact terms that we can drop, and we're left with 
 \eqn{\FT \left(\frac{1}{ x^{2n+\epsilon }}\right)=- \frac{(2\pi)^2 }{ 4^{n-1}(n-1)!\,(n-2)!}(-p^2)^{n-2}\ln p^2+\CO (\epsilon).}[FTlog]
Such $\ln (-s)$ terms, associated with dimensions that are integer or nearly integer,  are responsible for the discontinuities like \discXF.  The needed smallness of the anomalous dimensions, $\epsilon \ll 1$, fits with the optical theorem connection \Optical\ to the cross section, since that assumes production of weakly coupled final state particles.

\subsec{Superconformally-covariant operator product expansion}[SCOPE]
Superconformal theories have $Q_\alpha$, $\bar Q_{\dot \alpha}$, and $P_\mu$ as raising operators, generating the descendants.  The superconformal primaries are annihilated by the lowering operators, $S^\alpha$ and $\bar S^{\dot \alpha}$, and $K_\mu$ at the origin.    The algebra, and our sign conventions, can be found in \rcite{Fortin:2011nq}.     To quote a few examples, the superconformal charges act on scalar superconformal primary operators as 
\eqn{[S^\alpha, \CO(x)]=i x\cdot \bar \sigma^{\dot \alpha \alpha} [\bar Q_{\dot \alpha}, \CO(x)] ,
}[Sact]
\eqn{S^\beta Q_\alpha (\CO (x))=2(\sigma^{\mu\nu\phantom{\!\alpha}\!\beta}_{\phantom{\mu\nu}\!\alpha}x_{[\mu}\partial_{\nu]}+\delta_\alpha^{\phantom{\alpha}\!\beta}x\cdot\partial)\CO(x)-ix\cdot \bar \sigma^{\dot \alpha \beta}Q_\alpha \bar Q_{\dot \alpha} (\CO(x))+(2\Delta _\CO +3r_\CO)\delta_\alpha^{\phantom{\alpha}\!\beta}\CO(x), }[SQO]
where we define $S^\beta Q_\alpha (\CO (x))\equiv \{S^\beta ,[Q_\alpha, \CO (x)]\}$ etc.

Conserved currents are descendants of superconformal primary operators $J$ with $\Delta _J=2$ and $Q^2(J)=\bar Q^2(J)=0$,  \eqn{
j_\alpha(x) =Q_\alpha(J(x)),\qquad 
j_\mu(x) =-\tfrac{1}{4}\Xi _\mu (J(x)),
}[Jdesc]
where $\Xi _\mu \equiv \bar{\sigma}_\mu^{\dot{\alpha}\alpha}[Q_\alpha ,\bar{Q}_{\dot{\alpha}}]$. In superspace, 
\eqn{\CJ(z)=J(x)+i\theta j(x)-i\bar\theta \bar\jmath(x) -\theta \sigma ^\mu \bar \theta j_\mu(x) +\cdots,}[Jzis]
where $\cdots$ are derivative terms, following from the conservation equations $D^2\CJ =\bar D^2 \CJ =0$.
The superconformal supercharges act on $J(x)$ as in \Sact
\eqn{S^\alpha(J(x))=ix\cdot \bar \sigma^{\dot \alpha \alpha}\bar Q_{\dot \alpha}(J(x)), \qquad \bar S^{\dot \alpha}(J(x))=-ix\cdot \bar\sigma ^{\dot \alpha \alpha}Q_\alpha(J(x)),}[SJ]
vanishing at the origin.  Acting on the descendants as in \SQO with  $\Delta _J=2$ and $r_J=0$, 
\twoseqn{S^\alpha (j_\alpha(x))&=-ix\cdot \bar \sigma^{\dot \alpha \alpha}Q_\alpha \bar Q_{\dot \alpha}(J(x))+4(x\cdot\partial+2)J(x),}[Sj]{S^\alpha(j^\mu(x))&=3\bar{\sigma}^{\mu\dot{\alpha}\alpha}\bar{\jmath}_{\dot{\alpha}}(x)-2x\cdot\bar{\sigma}^{\dot{\alpha}\alpha}\bar{\sigma}^{\mu\nu\dot{\beta}}_{\phantom{\mu\nu\dot{\beta}}\!\dot{\alpha}}\partial_\nu\bar{\jmath}_{\dot{\beta}}(x).}[Sjmu]

The OPE of all the descendants \Jdesc\ follow from that of the primary operators,
\eqn{ J(x)J(0) =\sum _{\substack{\rm{sprimary}\\ \CO ^{(\ell)}}}\frac{c_{JJ}^{\CO ^{(\ell) }}}{ (x^2)^{\frac{1}{2}(4-\Delta _\CO )}}F_{JJ}^{\Delta _{\CO}}(x,P, Q, \bar Q)_{\mu _1\dots \mu _\ell}\CO ^{\mu _1\dots \mu _\ell}(0),}[JJsOPE]
where ``sprimary'' is shorthand for ``superconformal primary''.  As discussed in  \rcite{Fortin:2011nq}, current conservation $Q^2(J)=\bar Q^2(J)=0$ plays an important role in relating superconformal primary and descendant OPE coefficients.  Applying  the above relations to the LHS of \JJsOPE\  gives e.g.\ \rcite{Fortin:2011nq} (see also there for discussion about the sign)
\eqn{S^\alpha (J(x)J(0))=S^\alpha (J(x))J(0)=-ix\cdot \bar \sigma^{\dot \alpha \alpha}\bar \jmath_{\dot \alpha} (x) J(0).}[SJJ]
\eqn{j^\alpha (x)j_\alpha (0)=\tfrac{1}{2} Q^2(J(x)J(0)).}[QQJJ]
In SCFTs, the latter can also be written via 
\eqn{j_\alpha (x) j_\beta (0)=\frac{1}{ x^2}Q_\beta (ix\cdot \sigma \bar S)_\alpha (J(x)J(0))}[QSJ]
various such relations were noted in  \rcite{Fortin:2011nq}; just to quote a couple more,
\eqn{S^\alpha S^\beta (J(x)J(0))=\bar S^{\dot \alpha}\bar S^{\dot \beta}(J(x)J(0))=0,}[SSJ]
\eqn{
j_\mu(x)J(0) =\frac{x^2\eta_{\mu\nu}-2x_\mu x_\nu}{4x^4}\left[S\sigma^\nu\bar{S}-\bar{S}\bar{\sigma}^\nu S\right](J(x)J(0)).
}[JmuJ]
The RHS of the OPE is constrained by \SSJ\ and analogous relations in \rcite{Fortin:2011nq}, including the constraints from the generators of special conformal transformations.  These relate different OPE coefficients inside the $J(x)J(0)$ OPE in supersymmetric theories, yielding the full OPE in terms of the OPE coefficients for the superconformal primaries.  

As we showed in \rcite{Fortin:2011nq}, these constraints can be efficiently implemented in superspace, using the general formalism of \rcite{Osborn:1998qu}.  The only operators that can appear on the RHS of the $J(x)J(0)$ are real, $U(1)_R$ charge zero operators, with the superspace expansion ($\xi _\mu \equiv \theta \sigma _\mu \bar \theta$ and $\cdots$ are operators with non-zero R-charge)
\eqn{\CO ^{\mu _1\dots \mu _\ell}(x, \theta, \bar\theta)=A^{\mu _1\dots \mu _\ell}(x)+\xi _\mu B^{\mu \mu _1\dots \mu _\ell}(x)+\xi ^2 D^{\mu _1\dots \mu _\ell}(x)+\cdots.}[realo]
This is similar to the chiral-antichiral $\Phi \bar \Phi$ OPE considered in \rcite{Poland:2010wg}, and as there $\Xi ^\mu \equiv \bar \sigma ^{\mu \dot \alpha \alpha}[Q_\alpha, \bar Q_{\dot \alpha}]$, and $B^{\mu \mu _1\dots \mu _{\ell}}=-\frac{1}{4}\Xi ^\mu A^{\mu _1\dots \mu _\ell}$ and $D^{\mu _1\dots \mu _\ell}=-\frac{1}{64}\Xi _\mu B^{\mu \mu _1\dots \mu _\ell}-\frac{1}{16}\partial ^2 A^{\mu _1\dots \mu _\ell}$.  Operators $B_{\ell +1}$ in \realo\ decompose into Lorentz irreducible  representations 
$M_{\ell +1}$ of spin $\ell +1$, $N_{\ell -1}$ of spin $\ell -1$, and $L_\pm$ in the  $(\half \ell \pm \half, \half \ell \mp \half)$ representation of $SU(2)\times SU(2)$.   

Operators \realo\ with odd spin $\ell$ are odd under exchanging currents $J_a\leftrightarrow J_b$ in the OPE $J_a(x) J_b(0)$, and thus only appear, proportional to the structure constants $f_{abc}$, in non-Abelian theories.  Since for simplicity we consider the $J(x)J(0)$  for $G=U(1)$, only even-$\ell$ operators appear in the primary $J(x)J(0)$ OPE.  For operators \realo\ with $\ell$ even, this means that only the $A_{\ell\text{ even}}$ and $D_{\ell\text{ even}}$ operators contribute.  For operators \realo\ with $\ell$ odd, the $B _{\ell +1}\to  M_{\ell+1}, \ N_{\ell -1}$ components contribute.    

Superconformal symmetry and current conservation fully determine all current-current OPE superconformal descendant coefficients from those of the superconformal primaries, since as shown in \rcite{Fortin:2011nq},  the superspace dependence (in $z_i=(x^\mu, \theta _\alpha, \bar \theta _{\dot \alpha})_i$) of the associated three-point functions is fully determined:
\eqn{\vev{\CJ(z_1)\CJ(z_2) \CO ^{\mu _1\dots \mu _\ell}(z_3)}=\frac{1}{\osbx{\bar 13}{2}\osbx{\bar 3 1}{2}\osbx{\bar 23}{2}\osbx{\bar 3 2}{2}}t_{\CJ\CJ \CO _\ell }^{\mu _1\dots \mu _\ell}(X_3, \Theta _3, \bar \Theta _3).}[JJksope]
Current conservation implies that, for \realo\ of spin $\ell$ even or odd, respectively,  
\eqn{t_{\CJ\CJ \CO_{\ell \text{ even}}}^{(\mu _1\dots \mu _\ell)} =c_{JJ\CO _{\ell}}\frac{X_+^{(\mu _1}\cdots X_+^{\mu _\ell)}}{(X \cdot \bar X)^{2- \frac12(\Delta -\ell)}}\left[1-\frac{1}{4}(\Delta-\ell-4)(\Delta+\ell-6)\frac{\Theta ^2\bar \Theta ^2}{X \cdot \bar X}\right]-\text{traces},}[JJspine]
\eqn{t_{\CJ\CJ \CO_{\ell \text{ odd}}}^{(\mu _1\dots \mu _\ell)} =c_{JJ\CO _{\ell}}\frac{X_+^{(\mu _1}\cdots X_+^{\mu _{\ell-1}}}{(X \cdot \bar X)^{2-\frac12(\Delta -\ell)}}\left[X_-^{\mu _\ell)} -\frac{\ell(\Delta -\ell-4)}{\Delta -2}\frac{(X_-\cdot X_+)X_+^{\mu _\ell)}}{X \cdot \bar X}\right]-\text{traces}}[JJspino]
 in terms of the spin $\ell$ and dimension $\Delta \equiv \Delta _{\CO}\equiv \Delta _A$ of the operator $\CO$;  see  \rcite{Fortin:2011nq} and \rcite{Osborn:1998qu} for explanation about the notation.  The primary OPE coefficient fixes the coefficient $c_{JJ\CO _{\ell}}$ above, and then all descendant OPE coefficients are fully determined by the requirement that they reproduce \JJksope, \JJspine, \JJspino.  The superconformal relations are exhibited by expanding these expressions out in superspace.  For example, setting $\theta _{i=1,2}=\bar \theta _{i=1,2}=0$, these imply   \rcite{Fortin:2011nq} that the coefficients $c_{ijk}$ of the three-point functions satisfy
\eqna{c_{JJD_{\ell;\text{prim}}}&=-\frac{\Delta (\Delta +\ell)(\Delta -\ell -2)}{8(\Delta -1)}c_{JJ A_\ell},\cr
c_{JJN_{\ell-1}}&=-\frac{(\ell+2)(\Delta-\ell-2)}{\ell (\Delta+\ell)} c_{JJ M_{\ell +1}}.}[JJrelns]
The OPE coefficients $c_{ij}^k$ are related to the three-point coefficients $c_{ijk}$ by $c_{ij}^k=c^{k\ell}c_{ij\ell}$, where $c_{ij}$ are the two-point function coefficients, and then \JJrelns\ implies that
\eqna{c_{JJ}^{D_{\ell;\text{prim}}}&=-\frac{(\Delta -1)}{2 \Delta  (\Delta +\ell +1)(\Delta -\ell -1)}c_{JJ} ^{A_\ell},\cr
c_{JJ}^{N_{\ell-1}}&=-\frac{\ell (\ell+2)(\Delta+\ell+1)}{(\ell +1)^2 (\Delta-\ell -1)} c_{JJ}^{ M_{\ell +1}}.}[JJOPErelns]

\newsec{Implications of superconformally covariant OPE for General Gauge Mediation}[GGM]

The GGM  \rcite{Meade:2008wd} framework relates visible-sector soft SUSY breaking parameters to hidden sector current two-point functions (defined following the convention of \rcite{Dumitrescu:2010ha})\footnote{The last relation can be altered for spontaneously broken non-Abelian groups to $\vev{j_\mu^A (p)J^B(-p)}=ip_\mu f^{ABC}\vev{J^C}/p^2$, but Lorentz and gauge invariance imply that this doesn't contribute to the soft masses in any case.  See \rcite{Buican:2009vv, Intriligator:2010be} for discussion of GGM in such cases.}
\eqna{
\<J(x)J(0)\> & =C_0(x) \xrightarrow{\FT}\widetilde C_0(p),\\
\<j_{\alpha}(x)\bar{\jmath}_{\dot\alpha}(0)\> & =-i\sigma^\mu_{\alpha\dot{\alpha}}\partial_\mu C_{1/2}(x)\xrightarrow{\FT}\sigma ^\mu _{\alpha \dot \alpha}p_\mu \widetilde C_{1/2}(p),\\
\<j_\mu(x)j_\nu(0)\> & =(\eta_{\mu\nu}\partial^2-\partial_\mu\partial_\nu)C_1(x) \xrightarrow{\FT}-(\eta_{\mu\nu}p^2-p_\mu p_\nu)\widetilde C_1(p) ,\\
\<j_\alpha(x)j_\beta(0)\> & =\epsilon_{\alpha\beta}B_{1/2}(x)\xrightarrow{\FT}\epsilon_{\alpha\beta}\widetilde B_{1/2}(p),\\
\<j_\mu(x)J(0)\> & =0.
}[correls]
The functions $C_a(x)$ are  real and $B_{1/2}(x)$ is complex, though in potentially realistic models it must be possible to rotate it to be real, to avoid large CP violating phases.   If the theory were supersymmetric, all $C_a(x)$ would be equal, and $B_{1/2}(x)$ would be zero.  

The leading contribution to the above functions in the UV limit comes from the unit operator on the RHS of the  $J(x)J(0)$ OPE \jjex, 
\eqn{\hbox{UV limit}: C_a(x)=\frac{\tau }{ 16\pi ^4x^4} +\CO \left(\frac{1}{x^2}\right)\xrightarrow{\FT}  \widetilde C_a(p)=\frac{\tau }{ 16\pi ^2}\ln \frac{\Lambda^2}{ p^2}+\CO \left(\frac{1}{ p^2}\right).}[UV]
 The $C_a(x)$ all coincide at this order, as seen from the OPE and $Q(\mathds{1})=0$ \rcite{Meade:2008wd, Buican:2008ws}.
If the theory were exactly superconformal, only the unit operator could have an expectation value and \UV\ would be the full answer.

Another application of the OPE in the UV limit was discussed in \rcite{Dumitrescu:2010ha}:  it follows from the relations 
\threeseqn{\vev{\bar Q^2Q^2(J(x)J(0))}&=-8\partial ^2(C_0(x)-4C_{1/2}(x)+3C_1(x)),}[cdiffo]
 {\bar \sigma _\mu ^{\dot \alpha \alpha}\vev{Q_\alpha \bar Q_{\dot \alpha}(j^\mu (x) J(0))}&=-6\partial ^2 (C_0(x)-2C_{1/2}(x)+C_1(x)),}[cdiffi]{\vev{Q_\alpha \bar Q_{\dot \alpha}(j^\alpha  (x)\bar \jmath^{\,\dot \alpha}(0))}&=2\partial ^2 (C_0(x)+2C_{1/2}(x)-3C_1(x)),}[cdiffii][cdiff]
and the OPE, that the difference of any two $\widetilde C_a(p)$ in the UV vanishes at least as rapidly as $1/p^4$ in any renormalizable theory.  For example  \rcite{Dumitrescu:2010ha}, using the OPE 
 \eqn{j^\mu (x)J(0)\sim \frac{x^\mu \CO (0)}{ x^{6-\Delta _\CO}}+\frac{V^\mu (0)}{ x^{5-\Delta _V}}+\cdots,}[jmuJope]
 where $\CO$ and $V^\mu$ are scalar and vector operators, Lorentz invariance implies that 
 only $V^\mu$  can contribute to \cdiffi, with $V^\mu$ a conformal primary so  unitarity requires $\Delta _V\geq 3$ (saturated by a conserved current).   This implies $\widetilde C_0(p)-2\widetilde C_{1/2}(p)+\widetilde C_1(p)\leq \CO(1/p^4, \ln (p^2)/p^4)$ for large $p$.  Likewise, using \cdiffo\ and \cdiffii, any two $\widetilde C_a(p)$ differ by at most  $\CO(1/p^4, \ln (p^2)/p^4)$ in the UV \rcite{Dumitrescu:2010ha}. 
 
\subsec{Constraints from (approximate) broken superconformal symmetry} 
 
We expect / conjecture that the GGM functions can be constrained by applying the current-current OPE, with the Wilson coefficients approximately constrained by approximate UV superconformal symmetry (up to RG running differences).  The IR effect of supersconformal symmetry breaking appears via the non-zero expectation values of the various operators on the RHS of the OPE, namely the operators  \realo\ and their descendants. 

By  Lorentz invariance, only scalar operators can have non-zero expectation values and translation invariance implies that $P_\mu \to 0$ in one-point functions.  So only scalar conformal primaries can have non-zero one-point functions.  Such operators can only come from the superconformal primary operators \realo\ with spin $\ell =0$ or $\ell =1$: the $A^{\ell =0}$ and $D_{\rm prim}^{\ell =0}$ components of $\ell =0$ scalar superconformal primaries $\CO ^{\ell =0}$ in \realo, or the $N_{\ell -1=0}$ component of a primary $\CO ^{(\ell =1)\mu}$.  Likewise, the operator $V^\mu (0)$ in \jmuJope\ can be the superconformal primary components $A^\mu$ or $D_{\rm prim}^\mu$ of a superconformal primary spin $\ell =1$ operator, or from the $M^\mu$ component of a spin $\ell =0$ superconformal primary operator \realo.

Consider first the GGM function $C_0(x)$, which is given by the expectation value of the 
$J(x)J(0)$ OPE.  The $\ell =0$ conformal primaries that can contribute on the RHS of the OPE \JJsOPE\  yield 
(using \JJOPErelns\ with $\ell =0$)
\eqna{C_0(x)&=\sum _{\CO} \frac{c_{JJ}^{\CO }}{ (x^2)^{\frac{1}{2}(4-\Delta _\CO )}}\left( \vev{ A_{\CO }}-\frac{x^2}{2\Delta _\CO  (\Delta _\CO +1)}\vev{D_{\CO; {\rm prim} }}\right)\\
&\quad+\sum _{\CO ^\mu}\frac{c_{JJ}^{N_{\CO ^\mu  }}}{ (x^2)^{\frac{1}{2}(3-\Delta _{\CO ^\mu}  )} }\vev{N_{\CO ^\mu}}  
}[JJterm]
 $\CO$ runs over the real superconformal primaries with $\ell =0$, and $\CO^\mu$ over those with $\ell =1$, and $N_{\CO ^\mu}$ is the $\ell =0$  conformal primary, superconformal descendant.  The $\vev{D_{\CO;{\rm prim}}}$ and $\vev{N_{\CO ^\mu}}$ expectation values are  SUSY-breaking parameters of the low-energy theory.   As in the discussion in  \rcite{Dumitrescu:2010ha}, two simplifying limits are the small SUSY-breaking parameters limit, and the low-energy, spurion limit.  

The functions $C_{1/2} (x)$, $C_1(x)$, and $B_{1/2}(x)$ can similarly be written by applying the OPE to the current two-point functions on the LHS of \correls.   All of these descendant current two-point functions are fully determined by the $J(x) J(0)$ primary OPE.  In terms of the superspace expressions following \JJksope, we simply need to extract the appropriate $\theta _{1,2}$, $\bar \theta _{1,2}$ term, to pick out the $J$ descendant via \Jzis.   So $C_{1/2} (x) $ is found by applying $\partial _{\theta _{1, \alpha}}\partial _{\bar \theta _{2, \dot \beta}}$ to both sides of \JJksope\ before setting $\theta _{i=1,2}=\bar \theta_{i=1,2}=0$, and $C_1(x)$ is found by extracting the $\theta _1\sigma ^\mu \bar \theta _1\theta _2\sigma ^\nu\bar\theta _2f(\theta _3, \bar\theta _3)$ terms from \JJksope.  These lead to expressions for $C_{1/2}(x)$ and $C_1(x)$ analogous to \JJterm, fully determining them in terms of the same coefficients in \JJterm, the $C_{JJ}^\CO$ and $C_{JJ}^{N_{\CO ^\mu}}$ OPE Wilson coefficients and the vacuum expectation values $\vev{A_\CO}$, $\vev{D_{\CO; {\rm prim} }}$, and $\vev{N_{\CO ^\mu}}$.  Likewise, for the case of $B_{1/2}(x)$, using \QQJJ\ gives
\eqna{B_{1/2}(x)=\sum _{\CO} \frac{c_{JJ}^{\CO }}{ (x^2)^{\frac{1}{2}(4-\Delta _\CO )}}\vev{(Q^2A)_{\CO; {\rm prim} }}.
}[jjbterm]

The SUSY-breaking differences of the $C_a(x)$ can also be analyzed via \cdiffo\ to \cdiffii, applying the OPE to the current-current operators on the LHS.  As an example, applying the OPE to the LHS of \cdiffo, the contributing terms are the $D_{\ell =0}$ terms on the RHS of the OPE, so using $Q^2\bar Q^2 (A_{\ell})=-128 D_{\ell ; \rm{prim}}+\text{descendants}$,
\eqn{\frac{1}{16} \partial ^2 (C_0(x)-4C_{1/2}(x)+3C_1(x))=\sum _{\CO_{\ell =0}}\frac{c_{JJ}^{\CO }}{ (x^2)^{\frac{1}{2}(4-\Delta _\CO )}} \vev{D_{\CO; \rm{prim}}}
}[QQQQcdiff]

We can similarly consider the difference of the $C_a$'s in \cdiffi, using the OPE \jmuJope.    The $j_\mu (x) J(0)$ superconformal descendant OPE is fully determined from the $J(x)J(0)$ superconformal primary OPE.  One way to obtain this is to note that the $j_\mu (x_1) J(x_2)$ OPE can be obtained from the superspace three-point functions  \JJksope\ results \JJspine\ and \JJspino, by taking the  $\theta _1\sigma ^\mu \bar \theta _1$ component to get $j_\mu (x_1)$ and $\theta _2=\bar \theta _2=0$ to get $J(x_2)$.   Alternatively, we can use \JmuJ\ to get the $j_\mu (x) J(0)$ OPE from the $J(x) J(0)$ OPE.  This gives the conformal primary operator $V^\mu$ in \jmuJope, that contributes to \cdiffi, in terms of the operators $A^\mu _{\CO _{\ell =1}}$, $D^\mu _{\CO _{\ell=1}}$ and  $N^\mu _{\CO _{\ell =2}}$ and $M^\mu _{\CO _{\ell =0}}$.  Acting with $\Xi ^\mu =\bar \sigma ^{\mu \dot \alpha \alpha}[Q_\alpha, \bar Q_{\dot \alpha}]$ to get the LHS of \cdiffi, this gives an expression very analogous to \QQQQcdiff, that relates $\partial ^2 (C_0(x)-2C_{1/2}(x)+C_1(x))$ to the superconformal primary OPE coefficients $c_{JJ}^{\CO_{\ell =0}}$ and $c_{JJ}^{N_{\CO ^\mu}}$, along with the $\vev{D_{\CO_{\ell =0}}}$ and $\vev{N_{\CO ^\mu}}$ SUSY-breaking expectation values. 
 
\newsec{Analyticity properties of the GGM functions \texorpdfstring{$\widetilde C_a(p)$}{Ca(p)} and \texorpdfstring{$\widetilde B_{1/2}(p)$}{B12(p)}}[analggm]
 
Analyticity properties of correlation functions encode a wealth of physical information (see  e.g.\ \rcite{Itzykson:1980rh, Pindor:2006aa}).   The functions $\widetilde C_a(p^2)$ and $\widetilde B_{1/2}(p^2)$ \correls, coming from the hidden sector, contribute to the visible gauge vector multiplet propagators  (see e.g.\ \rcite{Intriligator:2010be}), so analyticity properties of the GGM functions connect with that of the gauge field propagators. 
 The functions $\widetilde C_a(s)$ and $\widetilde B_{1/2}(s)$ are analytic in $s=-p^2$, aside from cuts on the positive, real-$s$ axis for $s$ sufficiently large to create on-shell hidden sector states.  
 The discontinuities of the imaginary part of the $\widetilde C_a$ across the cut is then related by the optical theorem to total cross sections for hidden sector pair production, as in \Optical. As in \integrate, analyticity implies that the full GGM functions $A(s)=\widetilde C_a (s), \widetilde B_{1/2}(s)$ can be reconstructed from integrating their discontinuities along all their cuts, labeled by $c$:
\eqn{A(s)=\frac{1}{2\pi i}\sum _{c={\rm cuts}}\int _{s_{0, c}} ^\infty ds'\frac{[\Disc A(s')]_c}{s'-s}=\sum _c \frac{1}{\pi}\int _{s_0}^\infty ds'\, \frac{\Ima A(s')|_c}{ s'-s},}[integratec]
where $s_{0,c}$ and $[\Disc A(s')]_c$ are the cut's endpoint and discontinuity, respectively.  The OPE can be used to approximate the contribution from the large $s'$ UV part of the cut integral.  
 
 Let's first consider $\widetilde C_0(p^2)$, which can have a cut when the scalar (auxiliary) component $D(p)$ of the gauge multiplet can couple to produce a pair of on-shell scalars, of masses $m_1$ and $m_2$.  The production cross section for this process is
 \sproduction
 \eqn{\sigma _{0\to 0+0}(s)=\frac{\lambda ^{1/2}(s, m_1, m_2)}{ 8\pi s^2}\left|{\cal M} _{0}\right|^2,}[sprodii]
  where $\lambda ^{1/2}(s, m_1, m_2)=2\sqrt{s}|\vec p\hspace{1pt}|$ is the kinematic factor \Lambdais\ and ${\cal M}_0\equiv {\cal M}_{0\to 0+0}$.   The optical theorem \Optical\ relates this to the discontinuity 
 \eqn{\Disc \widetilde C_0(s)=\sum \frac{2s}{ (4\pi  \alpha) ^2}\sigma _{0\to 0+0} (s)=\sum \frac{\lambda ^{1/2}(s, m_1, m_2)}{ 4\pi s}\left|\frac{{\cal M}_0}{ 4\pi \alpha }\right|^2,}[spinzerodisc]
 where the sum is over all all distinct pairs of particles that can be produced. 
  
 Now consider $\widetilde C_{1/2}(p^2)$, which can have a cut where the gaugino component $\lambda _\alpha (p)$ of the gauge multiplet can create an on-shell spin $0+\half$ pair of states, of masses $m_0$ and $m_f$ respectively.\footnote{For (partially) Higgsed gauge messengers \rcite{Buican:2009vv, Intriligator:2010be}, we can also have $\sigma _{\half\to 1+\half}$, and also $\sigma _{1\to 1+0}$.}  The  total integrated cross section $\sigma =\int \!\frac{d\sigma }{ d\Omega}\,d\Omega$, averaged over initial spins and summed over the final ones, is given by (where ${\cal M}_\half \equiv {\cal M}_{\half \to 0+\half}$)
  \eqn{\sigma _{\half \to 0+\half}=\frac{\lambda ^{1/2}(s, m_s, m_f)}{ 8\pi s^2}\half \left(1+\frac{m_{f}^2-m _s^2}{ s}\right)|{\cal M}_\half|^2.}[fprod] 
The additional kinematic factor of $\half (1+(m_f^2-m_s^2)/s)$ compared with \sprodii\ comes from the spin factor sums and angular integration (see e.g.\ eq.\ (5.13) in \rcite{Peskin:1995ev}).   The discontinuity of $\widetilde C_{1/2}(p^2)$ is related to this cross section by the optical theorem,
\eqn{\Disc \widetilde C_{1/2}(s)=\sum \frac{s}{ 8\pi ^2\alpha ^2}\sigma _{\half \to 0+\half} (s)=\sum \frac{\lambda ^{1/2}(s, m_s, m_f)}{ 8\pi s}\half \left(1+\frac{m_{f}^2-m_s^2}{ s}\right)\left|\frac{{\cal M}_\half}{ 4\pi \alpha }\right|^2.}[spinhalfdisc]

Likewise for spin 1, a massless intermediate vector boson can decay to either two massive scalars or two massive fermions. In either case, the final state is a  CP conjugate pair, of the same mass.  Accounting for the spin-kinematic factors, the total cross sections are
\twoseqn{\sigma _{1\to 0+0}&=\frac{\lambda ^{1/2}(s, m_s, m_s)}{ 8\pi s^2}\frac{1}{ 6}\left(1-\frac{4m_s^2}{ s}\right)|{\cal M}_{1\to 0+0}|^2,}[spinonetozero]{\sigma _{1\to \half +\half}&=\frac{\lambda ^{1/2}(s, m_f, m_f)}{ 8\pi s^2}\frac{2}{ 3}\left(1+\frac{2m_f^2}{ s}\right)|{\cal M}_{1\to \half +\half}|^2.}[spinonetohalf]
The optical theorem gives the discontinuity of $\widetilde C_{1}$ in terms of these as 
\eqn{\Disc \widetilde C_{1}(s)=\frac{s}{ 8\pi ^2\alpha ^2}\left(\sum \sigma _{1\to 0+0} (s)+\sum \sigma _{1\to \half+\half}(s) \right),}[spinonedisc]
with the sums over the various scalars and fermions that can be produced.   In all of these discontinuities, $\lambda ^{1/2}(s, m_1, m_2)$ implies a cut, from $s_0=(m_1+m_2)^2$ to infinity.  

In the limit of unbroken supersymmetry, the produced state is a massive supersymmetric chiral superfield with $m_1, m_2, m_f, m_s\to m_{\text{SUSY}}$, and ${\cal M}_0={\cal M}_\half, {\cal M}_{1\to 0+0}, {\cal M}_{1\to \half +\half}\to {\cal M}_{\text{SUSY}}$.  All of the above total cross sections and discontinuities indeed properly coincide in this limit,
\eqn{\sigma _{\text{tot}; 0, \half, 1}\to \sigma _{\text{SUSY}}=\frac{1}{ 8\pi s}\sqrt{1-\frac{4m_{\text{SUSY}}^2}{s}}\left|{\cal M} _{\text{SUSY}, 0}\right|^2,}[SUSYprod]
with discontinuity 
\eqn{\Disc \widetilde C_{0, \half, 1}\to \Disc \widetilde C_{\text{SUSY}}=\sum _{m_{\text{SUSY}}}
\frac{1}{ 4\pi }\sqrt{1-\frac{4m_{\text{SUSY}}^2}{s}}\left|\frac{{\cal M} _{\text{SUSY}, 0}}{4\pi \alpha}\right|^2
,}[SUSYdisc]
 Even if SUSY is broken, in the large-$s$ limit the $\widetilde C_a(p^2)$ must all coincide to at least order $1/s^2$ \rcite{Dumitrescu:2010ha}, so their discontinuities at large $s$ must also coincide to this order.  

Finally, we can consider the possible cuts of $\widetilde B_{1/2}(s)$.  Much as with $\widetilde C_{1/2}$, such cuts can arise when the gaugino can produce on shell states.  Because $\widetilde B_{1/2}(s)$ is a complex rather than real amplitude, its cuts generally can not be identified with a real, positive-definite cross section.  On the other hand, to avoid CP violating phases, it should be possible to rotate $\widetilde B_{1/2}$ to be real in physically realistic theories.  As we will illustrate in an example, the cut structure of $\widetilde C_{1/2}$ and $\widetilde B_{1/2}$ are essentially the same, except that cut pairs add up in  $\widetilde C_{1/2}$, while they subtract in $\widetilde B_{1/2}$.  This opposite sign the SUSY-violating amplitude $\widetilde B_{1/2}(s)$ leads to a partial cancellation that is needed to ensure that  $\widetilde B_{1/2}(s)$, and hence its discontinuity, properly vanishes at least as fast as $1/s^2$ for large $s$.

The above discussion implicitly assumed an IR free spectrum for the produced states.  In that case, the discontinuities mentioned above come from $\ln (-s)$ terms in the current correlator Wilson coefficients, as illustrated in the appendix.  Such $\ln (-s)$ terms arise, as in \FTlog, from the Fourier transform of OPE coefficients of operators with integral (or half-integral) dimension, $2\Delta _\CO \in \IZ$.  More generally, we could contemplate a (broken) interacting SCFT, with mass gap, with quantum corrections to the anomalous dimensions leading to non-integer $2\Delta _\CO$.  The spectral analysis in that case is then similar to that considered in the context of ``unparticles," see e.g. \rcite{Rajaraman:2008bc}---we won't discuss it further here.

 \subsec{Soft masses from the OPE and analyticity}[softggm]
 
The OPE leads to approximations for the GGM soft masses in \GGMsoft, which can be applied even in strongly interacting hidden-sector theories.  Using the last expression in \integratec\ and applying the OPE,  the soft SUSY-breaking parameters are approximated by
\eqna{
M_{\rm gaugino} & \approx\sum_k\frac{\alpha\Ima[s^{d_k/2}\,\tilde{c}_{JJ}^k(s)]}{2^{d_k-1}d_kM^{d_k}}\<Q^2(\mathcal{O}_k(0))\>,\\
m_{\rm sfermion}^2 & \approx4\pi\alpha Y\<J(x)\>-\sum_k\frac{\alpha^2c_2 \Ima[s^{d_k/2}\,\tilde{c}_{JJ}^k(s)]}{2^{d_k+1}\pi d_k^2M^{d_k}}\<\bar{Q}^2Q^2(\mathcal{O}_k(0))\>.
}[SoftApprox]
Here the classical scaling dimension $d_k$ is related to the quantum scaling dimension by $\Delta_k=d_k+\gamma_k$, and $\Ima[s^{d_k/2}\,\tilde{c}_{JJ}^k(s)]$ is independent of $s=-p^2$ by dimensional analysis.   

Let us sketch a few details in how the expressions in \SoftApprox\ are obtained, to highlight in particular some approximations.   Using \GGMsoft\ and \integratec,
\eqna{M_{\rm gaugino}&=\pi i\alpha \widetilde B_{1/2}(s=0)=\alpha \int _{s_{0,c}}ds'\,\frac{ \Ima[i\widetilde B_{1/2}(s')]}{s'}\\
&\approx \alpha \sum _k  \int _{s_{\rm susy}}^\infty  ds'\, \frac{\Ima [\tilde c_{JJ}^k (s')]}{s'}  \langle {Q^2(\CO _k(0))\rangle}\\
&=\alpha\sum _k  \Ima [s^{d _k/2}\tilde c_{JJ}^k (s)] \langle {Q^2(\CO _k(0))\rangle} \int _{s_{\rm susy}}^\infty ds'\, (s' )^{-d _k/2 -1}.}[steps]
The second line of \steps\ involves two approximations.  First, we approximate $\widetilde B_{1/2}(s')$ by replacing it with its OPE---this is a good approximation for the large $s'$ part of the integral, while we apply it to the entire $s'$ integral.   

The next approximation is that the cut endpoints $s_{0,c}$ on the top line of \steps\ depend on the masses of the produced states, which are affected by the SUSY-breaking contributions, while on the second line we approximated all cuts as starting at the  unbroken-supersymmetric physical threshold $s_{\text{SUSY}}=4M^2$, where the SUSY-breaking corrections to the masses are dropped.  This is needed because, once we apply the OPE, the individual cuts are no longer visible.   While this approximation sounds perhaps rather crude, we will see in the example of weakly coupled messengers that it nevertheless gives the full answer, perhaps because the different individual cut locations essentially average to the supersymmetric threshold.  We replace $\Delta _k$ with the classical dimension $d_k$ to get the contributing $\ln (-s)$ contribution to the imaginary part in \steps.  Doing the $s'$ integral in \steps\ gives $M_{\rm gaugino}$ in \SoftApprox.  The derivation of $m^2_{\rm sfermion}$ is similar.  Uniform convergence is assumed, and the $m_{\rm sfermion}^2$ momentum integral was regulated to tame the otherwise IR-divergent integral.\footnote{Though the momentum integral is actually not IR-divergent but this cannot be inferred from the OPE alone; a complete knowledge of the $C_a(x^2M^2)$ functions is necessary. Also, although the OPE is convergent for large enough $s'$,  the approximations \SoftApprox might suffer from convergence issues from our integrating $s'$ all the way down to $s_{SUSY}=4M^2$.  This can require that the OPE sum be regulated by analytic continuation; an example of this will be seen in the next section.}  Notice that \SoftApprox only require the knowledge of the $J(x)J(0)$ OPE, which is constrained by OPE superconformality.

The expressions \SoftApprox\ can be further approximated by keeping only the contribution from the lowest dimension operator $\CO _K$ on the RHS of the OPE \jjex\ for which $Q^2(\CO _K)\neq 0$:
\eqna{
M_{\rm gaugino} & \approx-\frac{\alpha\pi w\gamma_{Ki}}{8M^2}\<Q^2(\mathcal{O}_i(0))\>,\\
m_{\rm sfermion}^2 & \approx4\pi\alpha Y\<J(x)\>+\frac{\alpha^2c_2w\gamma_{Ki}}{64M^2}\<\bar{Q}^2Q^2(\mathcal{O}_i(0))\>,
}[SoftApproxLow]
where $\gamma_{Ki}$ is the anomalous-dimension matrix which mixes $\CO _K$ with the operator $\mathcal{O}_i$.

%%%%%%%%%%%%%%%%%%%%%%%%%%%%%%%%%%%%%%%%%%%%%%%%%%%%%%%%%%%%%%%%%%%%%%%%%%%%%%%%%%%%%%%%%%%%%%%%%%%%%%%%%%%%%%%%%%%%%%%%%%%%%%%%%%%%%%%%%%%%%%%%%%%%

\newsec{Example: Minimal Gauge Mediation}[MGM]

We now apply and test our general ideas and methods in the canonical example of weakly coupled minimal gauge messenger mediation.  The theory  has canonical K\"{a}hler potential and a hidden-sector supersymmetry-breaking chiral superfield $X$ (or spurion) coupled to a pair of messengers $\Phi$ and $\widetilde{\Phi}$, of $U(1)$ charge $\pm 1$, via the superpotential
\eqn{
W_{\text{h}\otimes \text{m}}=h X\Phi\widetilde{\Phi}.}[SupPoti]
$X$ is chiral, $\bar{Q}_{\dot{\alpha}}(X(x))=0$, with $X(z_+)=X(y)+\sqrt2 \theta \chi (y)+\theta ^2 F(y)$, with \eqn{
\chi_\alpha(x) =\frac{i}{\sqrt{2}}Q_\alpha(X(x)),\qquad F(x)  =\frac{1}{4}Q^2(X(x)).
}[QX]
At low-energy, $X$ and $F$ get expectation values and the messengers $\Phi$ and $\widetilde \Phi$ become free fields with SUSY-split masses
\eqn{M_0=h \vev{X}, \qquad m_\pm^2 = m_0^2 \pm f,}[messmass]
with $M_0$ the fermion and $m_\pm$ the real-scalar masses ($m_0=|M_0|$ and $f=|h\vev{F}|$).  In the UV, with $X$ regarded as a dynamical field, the coupling $h$ in \SupPoti\ has a Landau pole; we restrict our attention to below the scale where it is UV completed or cutoff.

The $U(1)$ current superfield is $\CJ = \Phi ^\dagger \Phi -\widetilde \Phi ^\dagger \widetilde \Phi$, with $Q^2(J)=\bar Q^2(J)=0$ and components
\eqna{
J(x)&=\phi^\dagger\phi(x)-\tilde{\phi}^\dagger\tilde{\phi}(x),\\
j_\alpha(x)&=-i\sqrt{2}[\phi^\dagger\psi_\alpha(x)-\tilde{\phi}^\dagger\tilde{\psi}_\alpha(x)],\\
{\bar{\jmath}}_{\dot{\alpha}}(x)&=i\sqrt{2}[\phi\bar{\psi}_{\dot{\alpha}}(x)-\tilde{\phi}{\bar{\tilde{\psi}}}_{\dot{\alpha}}(x)],\\
j_\mu(x)&=i[\phi\partial_\mu\phi^\dagger(x)-\phi^\dagger\partial_\mu\phi(x)-\tilde{\phi}\partial_\mu\tilde{\phi}^\dagger(x)+\tilde{\phi}^\dagger\partial_\mu\tilde{\phi}(x)]+\psi\sigma_\mu\bar{\psi}(x)-\tilde{\psi}\sigma_\mu\bar{\tilde{\psi}}(x),
}[currents]
and their interactions with the SUSY-breaking superfield $X$ are given by
\eqn{
\mathscr{L}_\text{int}=-h^\ast h X^\dagger X(\phi^\dagger\phi+\tilde{\phi}^\dagger\tilde{\phi})-[h(-F\phi\tilde{\phi}+X\psi\tilde{\psi}+\phi\tilde{\psi}\chi+\tilde{\phi}\psi\chi)+\text{h.c.}].}[lint]
 We also define real superfields $K$ and $K'$, and ``meson" chiral field $M$, by
 \eqn{K=\Phi ^\dagger \Phi +\widetilde \Phi ^\dagger \widetilde \Phi, \qquad K'=K-2X^\dagger X, \qquad M=\Phi \widetilde \Phi.}[KKL]
 So in \lint\ $-|h X|^2$ sources the bottom component of $K$, and $hF$ sources $M$.  
 $K$ is the messenger's classical K\"ahler potential, with the classical dimension of a conserved current, but the current is violated by \SupPoti (though \SupPoti\ preserves $K'$):
 \eqn{Q^2 (K)=\frac{1}{8\pi ^2} W^2+4 hX M, \qquad Q^2(K')=\frac{1}{8\pi ^2} W^2.}
We include the anomaly term $W^2$ for completeness here, but it will not play a role in what follows since we initially turn off the gauge interactions, $\alpha \to 0$.  In this limit, $K'$ is a conserved current.

Below the scale of $\vev{X}$ and $\vev{F}$, where the theory is free, we know e.g.\
\eqna{\vev{J(x)J(0)}&\equiv C_0(x)=\frac{2}{(2\pi)^{d/2}}\left(\frac{m_+m_-}{x^2}\right)^{d/2-1}K_{d/2-1}(m_+x)K_{d/2-1}(m_-x),\\
\widetilde{C}_0(p^2)&=2\int \frac{d^4q}{(2\pi)^4}\frac{1}{q^2+m^2_+}\frac{1}{(p+q)^2+m_-^2}.
}[Czerois]
In the first line we used the $d$-dimensional propagator, with $K_\nu(z)$ a Bessel function.  In the following subsections we will test our general considerations by using the explicit, known expressions for the GGM functions  $\widetilde C_a(p^2)$ and $\widetilde B_{1/2}(p^2)$ in this case \rcite{Meade:2008wd}.  We will reinterpret the expressions in terms of the OPE in the ``CFT" \SupPoti\ with field $X$ included, applying and testing our constraints from superconformal symmetry.  Using e.g.\ \SQO, the superconformal supercharges act on the superfield $X$ components at $x^\mu =0$  as
\eqn{
\{S^\alpha, \chi_\beta(0)\} =3i\sqrt{2}r_X\delta^{\!\phantom{\beta}\alpha}_\beta X(0),\qquad
[S^\alpha,F(0)] =i\sqrt{2}(3r_X-2)\chi^\alpha(0), }
where $r_X=\frac{2}{3}\Delta _X$ is the R-charge of the chiral superfield $X$.  The $S^\alpha$ action at an arbitrary point $x$ is easily obtained from the superconformal-algebra equations and the chiral-superfield commutation relations.

\subsec{The cross sections and analyticity properties}[Analyticity]
The total cross sections for scattering from the visible to the hidden sector can be immediately computed to ${\cal O}(\alpha)$ from the general expressions \sprodii, \fprod, and \spinonetozero\ and \spinonetohalf.   In this weakly coupled hidden sector, the amplitude in these expressions is simply ${\cal M}=4\pi \alpha$, with the kinematic factors involving the hidden-sector messenger masses \messmass:
\eqna{
\sigma_0(\text{vis}\rightarrow \text{hid})&=\frac{(4\pi \alpha )^2}{4\pi s}\frac{1}{2s}\lambda^{1/2}(s,m_+,m_-),\\
\sigma_{1/2}(\text{vis}\rightarrow \text{hid})&=\frac{(4\pi \alpha )^2}{4\pi s}\frac{1}{4s^2}\left[(s+m_0^2-m_+^2)\lambda^{1/2}(s,m_0,m_+)+(m_+\rightarrow m_-)\right],\\
\sigma_{1}(\text{vis}\rightarrow \text{hid})&=\frac{(4\pi \alpha)^2}{4\pi s}\frac{1}{12 s^2}\left[(s-4m_+^2)\lambda^{1/2}(s,m_+,m_+)+(m_+\rightarrow m_-)\right.\\
&\left.\hspace{6cm}+4(s+2m_0^2)\lambda^{1/2}(s,m_0,m_0)\right],\\
\sigma^\prime_{1/2}(\text{vis}\rightarrow \text{hid})&=\frac{(4\pi \alpha)^2}{4\pi s}\frac{1}{2s}\left[\lambda^{1/2}(s,m_0,m_+)-\lambda^{1/2}(s,m_0,m_-)\right].
}[CrossSections]
Here  $\sigma^\prime_{1/2}$ is not an honest cross section, but we anyway relate it to $\widetilde B_{1/2}$, whose phase can be eliminated to make  $\sigma^\prime_{1/2}$ real and positive.  In the unbroken-SUSY limit, $F\to 0$,
\eqn{\sigma _{a=0, 1/2, 1}(s) \to \sigma _{\text{SUSY}}(s, m_{\text{SUSY}})=\frac{(4\pi \alpha )^2}{8\pi s}\sqrt{1-\frac{4m_{\text{SUSY}}^2}{s}}\theta (s-4m_{\text{SUSY}}^2), \qquad \sigma ' _{1/2}\to 0.}[susycross]
The full cross sections \CrossSections\ can be obtained from $\sigma_\text{SUSY}$ \susycross, e.g.\
\eqn{
\sigma_0(s)=\exp\left(-\frac{f^2}{s}\frac{\partial}{\partial m_0^2}\right)\sigma_{\text{SUSY}}(s),
}
with similar (but slightly uglier) expressions for $\sigma _{1/2}$, $\sigma _1$, and $\sigma ^\prime _{1/2}$.

The cross sections \CrossSections\ have expansions in  powers of $1/s$ in the UV limit, using  \messmass, 
\eqna{
\sigma_0(\text{vis}\rightarrow \text{hid})&=\frac{(4\pi \alpha)^2}{4\pi s}\left[\frac{1}{2}-\frac{m_0^2}{s}+\frac{f^2-m_0^4}{s^2}-\frac{2m_0^2 (m_0^4-f^2)}{s^3}+\mathcal{O}(s^{-4})\right],\\
\sigma_{1/2}(\text{vis}\rightarrow \text{hid})&=\frac{(4\pi \alpha)^2}{4\pi s}\left[\frac{1}{2}-\frac{m_0^2}{s}+\frac{\half f^2-m_0^4}{s^2}-\frac{2m_0^6}{s^3}+\mathcal{O}(s^{-4})\right],\\
\sigma_{1}(\text{vis}\rightarrow \text{hid})&=\frac{(4\pi \alpha)^2}{4\pi s}\left[\frac{1}{2}-\frac{m_0^2}{s}+\frac{f^2-m_0^4}{s^2}-\frac{2m_0^2 (m_0^4-f^2)}{s^3}+\mathcal{O}(s^{-4})\right],\\
\sigma_0-4\sigma_{1/2}+3\sigma_1&=\frac{(4\pi \alpha)^2}{4\pi s}\frac{2f^2}{s^2}\left[1+\frac{4m_0^2}{s}+\mathcal{O}(s^{-3})\right],\\
\sigma^\prime_{1/2}(\text{vis}\rightarrow \text{hid})&=-\frac{(4\pi \alpha)^2}{4\pi s}\frac{f}{s}\left[1+\frac{2m_0^2}{s}+\frac{6m_0^4}{s^2}+\mathcal{O}(s^{-3})\right].
}[snoOPE]
In the UV limit, the SUSY-breaking differences of $\sigma _0$, $\sigma _{1/2}$, and $\sigma _1$ show up at $\CO (f^2/s^3)$, while $\sigma '_{1/2}$ is $\CO (f/s^2)$. 

The optical theorem relations \spinzerodisc, \spinhalfdisc, \spinonedisc, relate these cross sections to the discontinuities of the GGM functions $\widetilde C_a(s)$, and here $\widetilde B_{1/2}(s)$ obeys a similar relation,
\eqna{
\sigma _{a=0, 1/2, 1}&=\frac{(4\pi\alpha)^2}{s}\Ima(i\widetilde{C}_a(s))=\frac{(4\pi\alpha)^2}{s}\frac{1}{2i}\Disc(i\widetilde{C}_a(s)),\cr
\sigma '_{1/2}&=\frac{(4\pi\alpha)^2}{m_0s}\Ima(i\widetilde{B}_{1/2}(s))=\frac{(4\pi\alpha)^2}{m_0s}\frac{1}{2i}\Disc(i\widetilde{B}_{1/2}(s)).
}[GGMoptical]
We now verify these relations from the known, explicit integral expressions for the GGM functions in this case, as given in \rcite{Meade:2008wd}.    Let's first remark that since, as shown on general grounds in \rcite{Dumitrescu:2010ha}, the $\widetilde C_a(s)$ coincide to $\CO (1/s^2, \ln s/s^2)$ in the UV limit, it follows from \GGMoptical\ that the $\sigma _a$ in this limit necessarily always coincide to $\CO (1/s^3)$, as seen explicitly in the present example in \snoOPE.  

Consider first  $C_0$ using its integral expression
\eqn{
\widetilde{C}_0=2\int \frac{d^4q}{(2\pi)^4}\frac{1}{q^2+m^2_+}\frac{1}{(p+q)^2+m_-^2}\supset -\frac{i}{8\pi^2}\int_0^1 dx\, \ln[x(1-x)p^2+xm_+^2+(1-x)m_-^2],
}
where the last expression is the finite part. The Landau equations for determining the endpoint of the cut, 
\eqn{
x(1-x)p^2+xm_+^2+(1-x)m_-^2=0\quad \text{and} \quad \frac{\pd}{\pd x}[x(1-x)p^2+xm_+^2+(1-x)m_-^2]=0,
}
have solutions $s_\pm =-p^2_\pm =(m_+\pm m_-)^2$ and $x_\pm = \frac{m_-}{m_-\pm m_+}$.  The $s_+$, $x_+$ solution gives the endpoint of the cut, while $s_-$ is  unphysical, since it has $x_-<0$, outside of the region of integration.  Indeed,  $\Delta \widetilde{C}_0$ can be here be calculated analytically from the integral to give
\eqn{\widetilde{C}_0\supset \frac{i}{8\pi^2 s}\lambda^{1/2}(s,m_+^2,m_-^2)\ln\frac{\sqrt{-s+(m_++m_-)^2}+\sqrt{-s+(m_+-m_-)^2}}{\sqrt{-s+(m_++m_-)^2}-\sqrt{-s+(m_+-m_-)^2}}.}
The only physical branch point, on the first sheet of the logarithm, is that at $s_+$---there is no physical branch point at $s_-=(m_+-m_-)^2$, and there is no physical pole at $s=0$.  Thus, in agreement with the above cross section and the optical theorem \GGMoptical,
 \eqn{
\Disc \widetilde{C}_0\equiv\widetilde{C}_0(s+i\epsilon)-\widetilde{C}_0(s-i\epsilon)=\frac{\lambda^{1/2}(s,m_+^2,m_-^2)}{4\pi s}\theta(s-(m_++m_-)^2).
}[cod]

It similarly follows from the explicit integral expression for $\widetilde C_{1/2}(s)$, 
\eqn{
\widetilde{C}_{1/2}=-\frac{2}{p^2}\int \frac{d^4q}{(2\pi)^4}\left[\frac{1}{(p+q)^2+m_+^2}+\frac{1}{(p+q)^2+m_-^2}\right]\frac{p\cdot q}{q^2+m^2_0,}
}
that $\widetilde C_{1/2}$ has two (physical) branch points,  at 
$s=(m_0+m_+)^2$ and $s=(m_0+m_-)^2$, with 
\eqn{
\Disc \widetilde{C}_{1/2}=\frac{1}{8\pi s^2}(s+m_0^2-m_+^2)\lambda^{1/2}(s,m_0,m_+)\theta(s-(m_0+m_+)^2)+[m_+\to m_-].
}[chalfd]
(Again, $s=0$ is not a pole on the first sheet of the logarithm.)  So \chalfd\ indeed agrees with \GGMoptical\ and the above cross sections.  Similarly, $\widetilde{B}_{1/2}$, has two branch points, at exactly the same positions in the $s$-plane as $\widetilde{C}_{1/2}$, with 
\eqn{
\Disc \widetilde{B}_{1/2}=\frac{m_0}{4\pi s}\lambda^{1/2}(s,m_0^2,m_+^2)\theta(s-(m_0+m_+)^2)-(m_+\to m_-).
}[bhalfd]
The relative sign between the two terms in \bhalfd\ cancels the contributions to ${\cal O}(1/s)$, consistent with the restoration of supersymmetry in the deep UV. 

Similarly, the explicit expression for $\widetilde C_1$, 
\eqn{\widetilde{C}_1=\frac{2}{3p^2}\int \frac{d^4q}{(2\pi)^4}\left\{ \frac{(p+q)\cdot(3p+2q)+4m_+^2}{(q^2+m_+^2)[(p+q)^2+m_+^2]}+(m_+\rightarrow m_-)-\frac{4q\cdot(p+q)+8m_0^2}{(q^2+m_0^2)[(p+q)^2+m_0^2]}\right\}
} 
reveals three branch points, at $s=4m_{\pm}^2,4m_0^2$.  (The supertrace relation $\Str M^2=0$, i.e.\ $m_+^2+m_-^2-2m_0^2=0$, is needed to prevent $\widetilde C_1(s) $ from having a pole at $s=0$ on the physical sheet.)  The $\widetilde C_1(s)$ discontinuities are consistent with the optical theorem and the cross sections \spinonetozero\ for scalar production and \spinonetohalf\ for fermion production.   At large $s$, the sum of the discontinuities across the three cuts add to coincide with that found above for $\widetilde C_0$ and $\widetilde C_{1/2}$ to order $\CO(1/s^2)$, consistent with UV supersymmetry restoration. 

\subsec{OPE for \texorpdfstring{$J(x)J(0)$}{J(x)J(0)} and superpartners}

We now consider the current-current OPE $J(x)J(0)$ \jjex, along with its Fourier transform
\eqna{i\int d^4 x\, e^{-ip\cdot x}J(x)J(0)&\to\tilde c_\mathds{1}(s, \Lambda) \mathds{1}+ \tilde{c}_K(s) K(0)+\tilde{c}_{J^2}(s)J^2(0)+\tilde{c}_{K^2}(s)K^2(0)+\cdots\\
&\quad +{\cal F}(X, X^\dagger, F, F^\dagger, \chi , \chi ^\dagger; s, \mu).}[OPEJJi]
The first few terms in the position-space OPE are found from taking Wick contractions
\eqn{c_{\mathds{1}}(x) =\frac{1}{8\pi ^4 x^4}+\cdots,  \qquad c_K(x)=\frac{1}{2\pi ^2 x^2}+\cdots, \qquad c_{J^2}(x)=1+\cdots,}[posfew]
(So $\tau =w=2$ in \jjex, coming from $\Phi$ and $\widetilde \Phi$. In the $\alpha, h \to 0$ limit, $K$ becomes a conserved current and the leading $K$ term on the RHS of the $J(x) J(0)$ OPE can be regarded as giving the $\Tr U(1)_J^2 U(1)_K=2$ 't Hooft anomaly.)  Here  $\cdots$ are higher order perturbative corrections.  These Wilson coefficients have Fourier transforms
\eqn{\tilde{c}_{\mathds{1}}(s)=\frac{1}{8\pi ^2} \ln \frac{\Lambda ^2}{-s}+\cdots, \qquad \tilde c_K(s)=-\frac{2}{s}+\cdots, \qquad \tilde c_{J^2}(s)=\delta ^{(4)}(p)+\cdots .}[momfew]
For example, $\tilde c_K$ can be found from the diagram 
 \begin{figure}[H]
\centering
\includegraphics{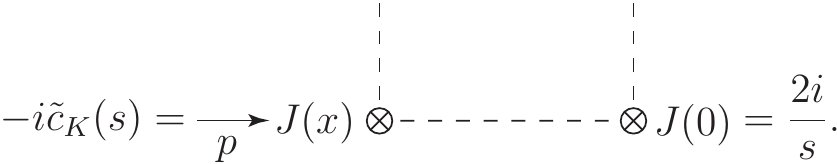}
%\pstool{./figures/cks}{\psfrag{L}[][]{$J(0)$}
%\psfrag{J}[][]{$J(x)$}
%\psfrag{p}[][]{$p$}
%\psfrag{c}[][]{$-i\tilde{c}_K(s)=$}
%\psfrag{r}[][]{$=\displaystyle\frac{2i}{s}.$}}
\end{figure}
\vspace{-2.05cm}
\eqn{\phantom{x}}[ClCoeffK]

\vspace{0.3cm}
\noindent
As usual, a  UV cutoff $\Lambda$ enters for the Fourier transformation of the identity term in \OPEJJi.   

The important terms in what follows will be those on the second line of \OPEJJi, representing the contributions of the supersymmetry breaking ``goldstino" (or spurion background) superfield $X$, and its superpartners, to the OPE.   When we take expectation values of \OPEJJi, and superpartners, the superconformal and supersymmetry breaking effects will come from the expectation values of these operators involving $X$ and $X^\dagger$.   Since $J(x)$ is $U(1)_R$ neutral, the possible terms in ${\cal F}$ in \OPEJJi\ include 
\eqn{\begin{split}
i\int d^4 x\, e^{-ip\cdot x}J(x)J(0)&\supset \sum_{m,n=0}^{\infty} \tilde{c}_0(m,n;s,\mu)(F^\dagger F)^m(X^\dagger X)^n(0)\\
 &\quad+ \sum_{m,n=0}^{\infty} \tilde{d}_0(m,n;s,\mu)(F^\dagger F)^m(X^\dagger X)^n X^\dagger F^\dagger\chi^2(0)+\text{h.c.}\\
 &\quad+ \sum _{m, n=0}^\infty \tilde e_0(m, n; s, \mu) (F^\dagger F)^m (X^\dagger X)^n \chi ^2\bar \chi ^2 (0)
 +\cdots.
\end{split}}[JJXterms]
There are similar OPE expansions for the current superdescendants of $J(x)$, e.g.\
\eqn{\begin{split}
i\int d^4 x\, e^{-ip\cdot x}j_\alpha (x)\bar \jmath_{\dot \alpha}(0)&\supset -i\sigma ^\mu _{\alpha \dot \alpha} p ^\mu \sum_{m,n=0}^{\infty} \tilde{c}_{1/2}(m,n;s,\mu)(F^\dagger F)^m(X^\dagger X)^n(0)\\
 &\quad-i\sigma ^\mu _{\alpha \dot \alpha }p ^\mu \sum_{m,n=0}^{\infty} \tilde{d}_{1/2}(m,n;s,\mu)(F^\dagger F)^m(X^\dagger X)^n X^\dagger F^\dagger\chi^2(0)+\text{h.c.} \\
 &\quad -i\sigma ^\mu _{\alpha \dot \alpha }p ^\mu \sum_{m,n=0}^{\infty} \tilde{e}_{1/2}(m,n;s,\mu)(F^\dagger F)^m(X^\dagger X)^n X^\dagger F^\dagger\chi^2\bar \chi ^2(0)
+  \cdots.
\end{split}}[JaXterms]

The scale $\mu$ appearing in \JJXterms\ is the IR normalization point mentioned in section \OPE.  The Feynman diagrams used to compute the Wilson coefficients in \JJXterms (see appendix \ref{AppWils}), are UV-convergent but IR-divergent.  So we integrate over virtual momenta starting at an IR cutoff $\mu$, yielding $\mu$ dependent Wilson coefficients that are governed by  the RG equations \RGe.  Operator expectation values are similarly $\mu$ dependent, governed by RG equations.  The $\mu$ dependence ultimately drops, as discussed in \rcite{Novikov:1984rf}, when computing OPE expectation values, like the GGM functions. This here works thanks to operator mixing between operators on the two lines  of \OPEJJi, involving the messengers and $X$. 

\begin{figure}[ht]
\centering
\includegraphics{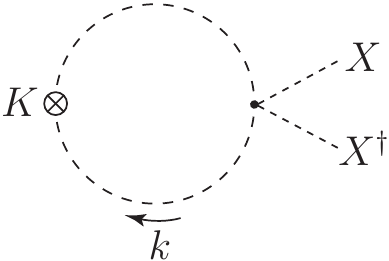}
%\pstool{./figures/OPEMixKXX}{\psfrag{K}[][]{$K$}
%\psfrag{k}[][]{$k$}
%\psfrag{X}[][]{$X$}
%\psfrag{S}[][]{$X^\dagger$}}
\vspace{-0.3cm}
\caption{Diagram giving rise to operator mixing between $K$ and $X^\dagger X$.}\label{FigKXX}
\end{figure}

As an example of this, consider the coefficient $\tilde c_{X^\dagger X}(s, \mu) $ of the operator $X^\dagger X$ in the OPE, called  $\tilde{c}_0(0,1;s,\mu)$ in \JJXterms, which is obtained at one-loop in the appendix by evaluating a Feynman diagram with an insertion of $X^\dagger X$, with IR cutoff $\mu$ on the loop momentum,
\eqn{\tilde c_{X^\dagger X}(s, \mu)=\frac{1}{4\pi ^2}\frac{|h |^2}{s}\ln \frac{-s}{\mu ^2} +\cdots,}[cXX]
where again $\cdots$ includes higher order corrections in $|h|^2$.  
The $\mu$ dependence in \cXX\ is cancelled, effectively  replaced with $\Lambda$, by  the one loop operator mixing between $X^\dagger X$ and the operator $K$, given by the diagram of Fig.\ \ref{FigKXX}.  

\noindent This diagram, which requires both UV cutoff $\Lambda$ and IR cutoff $\mu$, gives operator mixing:
\eqn{
K_{\text{ren}}(0)=K(0)-\frac{|h|^2}{8\pi^2}\ln\frac{\Lambda^2}{\mu^2}X^\dagger X(0). 
}[KXXMixing]
(This is related to the fact that $K$ in \KKL\ has $\gamma _K=|h|^2/16\pi ^2$ whereas $K'$ in \KKL\ has $\gamma _{K'}=0$ to this order.) 
When combined with the tree-level Wilson coefficient $\tilde c_K$ in \momfew,  the $\mu$-dependence in \cXX\ cancels with that in \KXXMixing, and 
is thereby ultimately replaced with a $\Lambda$-dependence from $\tilde{c}_{K}(s)K(0)$.  

 As an immediate illustration and check of our methods and results, let us connect the first few leading UV terms of the $J(x)J(0)$ OPE expectation value with the corresponding terms in the $\sigma _0(s)$ cross section in \snoOPE.  Using \GGMoptical,  \snoOPE, and \OPEJJi, we have
 \eqna{\Disc \widetilde C_0(s)&=\frac{1}{2\pi}\left[\frac{1}{2}-\frac{m_0^2}{s}+\frac{f^2-m_0^4}{s^2}-\frac{2m_0^2 (m_0^4-f^2)}{s^3}+\mathcal{O}(s^{-4})\right]\\
 &=\Disc \tilde c_{\mathds{1}}(s) \vev{\mathds{1}}+\Disc \tilde c_{X^\dagger X} (s)\vev{X^\dagger X}+\Disc \tilde c_{(X^\dagger X)^2}\vev{(X^\dagger X)^2}+\cdots}[coexp]
 The first two terms on the top line indeed agree with the first two terms on the second line, upon using $\tilde c_{\mathds{1}}(s)$ from \momfew, and $\tilde c_{X^\dagger X}(s)$ from \cXX\ and 
 \eqn{\vev{\mathds{1}}=1, \qquad \vev{(|h|^2X^\dagger X)^n}=m_0^{2n}, \qquad \Ima\ln(-(s\pm i\epsilon))=\mp \pi.}[vdisc]
 
 In  fact, we can reproduce the full cross sections \CrossSections\ and associated discontinuities, from the OPE \OPEJJi\ expectation value, 
 \eqn{\widetilde C_0 \supset \sum _{m, n=0}^\infty \tilde c_0(m, n; s) (\vev{F^\dagger F})^m(\vev{X^\dagger X})^n.}[JJvev]
 An explicit one-loop computation of the Wilson coefficients $\tilde c_0(m, n; s, \mu)$ in \JJXterms\ is given in the appendix.  The discontinuity in particular comes from the terms $\sim \ln (-s)$ as in \vdisc, and using the result from the appendix gives 
 \eqn{\Disc\tilde c_0(m, n; s) =-\frac{1}{2\pi}\frac{(-1)^{m}\Gamma(2(m+n)-1)}{\Gamma(m+n)\Gamma(m+1)\Gamma(n+1)}\left(\frac{1}{s}\right)^m\left(\frac{|h|^2}{s}\right)^{m+n}.}[codisc]
Using \codisc, the seemingly complicated series in $m$ and $n$ indeed nicely sums up to give (recall from \messmass\ that $m_0\equiv |h\vev{X}|$ and $f\equiv |h\vev{F}|$)
\eqna{\Disc\widetilde C_0&= \sum _{m, n=0}^\infty \Disc\tilde c_0(m, n; s) (\vev{F^\dagger F})^m(\vev{X^\dagger X})^n \\
&=\frac{1}{4\pi s}\sqrt{ s^2 -4 m_0^2 s+4 f^2}.}[Cdmatch]
Upon using \messmass, \Cdmatch\ indeed exactly reproduces, to all orders in $1/s$, the expression \cod, involving the standard kinetic factor $\lambda ^{1/2}(s, m_+, m_-)$ \Lambdais. 

As indicated in the $J(x)J(0)$ OPE \JJXterms, there are terms involving $X$'s fermion components, $\chi$ (the goldstino).  Such terms vanish upon taking the expectation value, so they do not contribute to $\widetilde C_0(s)$, as in \JJvev.   We retain the $\chi$ terms in \JJXterms\ because they do contribute once we act on them with the supercharges $Q$, $\bar Q$, so they contribute to \GGMsoft, \cdiff etc.  The form of the terms in \JJXterms\ have been constrained by the $U(1)_R$ symmetry and reality of $J$.\footnote{ There are additional operators involving derivatives, with OPE coefficient denoted as e.g. $\partial d_0$:
\eqn{\sum _{m, n}\partial d_0 (m, n; s, \mu) (F^\dagger F)^m (X^\dagger X)^n  \partial _\mu \chi \sigma ^\mu \bar \chi (0).}} The action of $Q$ on the operators in \JJXterms\ can be obtained from \QX, which we can represent as 
\eqn{Q_\alpha \to -i\sqrt{2}\left(\chi _\alpha {\partial \over \partial X}+F{\partial \over \partial \chi^\alpha}\right),}[QXi]
so e.g.\
\eqn{Q^2\to 4F\frac{\partial}{\partial X}-2\chi^2\frac{\partial^2}{\partial X^2}-4\chi^\alpha F \frac{\partial^2}{\partial X\,\partial\chi^\alpha}+2F^2\frac{\partial^2}{\partial\chi^2}.}[QXii]

Let us now consider the $j_\alpha (x)j_\beta (0)$ OPE, whose expectation value gives $B_{1/2}(x)$.  By  relation \QQJJ, this can be obtained from $Q^2$ acting on $J(x)J(0)$ OPE \JJXterms, and   the terms with non-zero expectation value are those without remaining $\chi$ or $\chi ^\dagger$ fermion  fields.  In terms of \QXii, the contributions come from the first and last terms, giving 
\eqn{
i\int d^4 x\,e^{-ip\cdot x}\vev{j_\alpha(x)j_\beta(0)}\to\epsilon_{\alpha\beta}\vev{FX^\dagger\sum_{m,n=0}^{\infty}\tilde{c}^{\,\prime}_{1/2}(m,n;s,\mu)(F^\dagger F)^m(X^\dagger X)^n},
}
with coefficients $\tilde c^{\,\prime} _{1/2}$ contributions from the $ \tilde c_0$ and $\tilde d_0$ terms in \JJXterms
\eqn{
\tilde{c}^{\,\prime}_{1/2}(m,n; s, \mu)=(n+1)\tilde{c}_0(m,n+1; s, \mu)+2\tilde{d}_0(m-1,n; s, \mu).}[CoeffRel]
Using the explicit expressions for $\tilde c_0(m, n; s, \mu)$ and $\tilde d_0(m, n; s, \mu)$, given by \eqref{WilCoeff} and \eqref{WilCoeffd} in the appendix, we find that \CoeffRel\ indeed gives the correct expression for $\widetilde B_{1/2}(s)$, and in particular its discontinuity is properly related to the last expression in \CrossSections\ and \snoOPE:
\eqna{\Disc\widetilde B_{1/2}(s)&=\sum _{m, n=0}^\infty \Disc\tilde c_{1/2}^{\,\prime}(m, n; s) (\vev{F^\dagger F})^m(\vev{X^\dagger X})^n \\
&=\frac{m_0}{4\pi s}\sqrt{ s^2 -4 m_0^2 s -2fs + f^2}-(f\to -f),}[Bdmatch]
which precisely reproduces \bhalfd. 

We can similarly consider $Q^2\bar Q^2$ acting on the $J(x)J(0)$ OPE, which by \cdiffo\ gives expectation value equal to $-8\partial ^2(C_0(x)-4C_{1/2}(x)+3C_1(x))$.  Now, using \QXii\ and its analog for $\bar Q^2$, the $\tilde c_0(s)$, $\tilde d_0(s) \chi ^2+\text{h.c.}$, and $\tilde e_0(s) \chi ^2\bar \chi ^2$ terms in the $J(x)J(0)$ OPE \JJXterms\ all contribute.  The resulting relation can be verified from a direct loop computation of the $\tilde e_0(s)$ Wilson coefficients, along the lines of the $\tilde c_0$ and $\tilde d_0$ perturbative computation outlined in the appendix. 

Let us now turn to using, and checking, the additional constraints that follow from our claimed 
 superconformal covariance of the OPE Wilson coefficients.  One way to implement the constraints of superconformal invariance is to directly use the superspace-based \rcite{Osborn:1998qu} results of  \rcite{Fortin:2011nq}, reviewed in section \SCOPE above.  It follows from these results that the 
OPE of all components of the $\CJ (z)$ current superfield \Jzis\ are fully determined by the superconformal primary contributions to the  primary $J(x)J(0)$ OPE, with independent Wilson coefficients for all real superconformal primary operators ${\cal O}^{\mu _1\dots \mu _\ell}$ \realo.  As discussed in \JJterm, only the spin $\ell =0$ operators ${\cal O}$, and spin $\ell =1$ operators ${\cal O}^\mu$ have spin zero components that can get expectation values and contribute to the GGM functions.  

To use these results here, we need to classify the independent, real superconformal primary operators of spin $\ell =0,1$ that can be built from $X$ and $X^\dagger$. 
Clearly one such class of primary operator superfields are $\CO _n(z)=(X^\dagger X)^n$.  Using 
\QXii\ $Q^2(X^n)= n X^{n-2}(4FX -2(n-1) \chi ^2)$, we see that the descendants in \realo\ involve particular linear combinations of $FX$ and $\chi ^2$.  Classes of additional superconformal primary operators can be obtained from different, orthogonal linear combinations of $FX$ and $\chi ^2$ terms.\footnote{It is necessary here to  retain the interaction \SupPoti, since $F$ is a null operator if $X$ is free.} We won't work out here the details of all classes of superconformal primaries for this example.

Alternatively, we can directly check that the superconformal relations like \SJJ, \QSJ, \SSJ, \JmuJ\ etc.\ are satisfied.  As a first example, conformal covariance with respect to $K_\mu$ fully determines the $P$ dependence in \OPEx (as in e.g.\ \rcite{Ferrara:1973yt}), and in particular the contribution of scalar operators $\CO$ to the $J(x)J(0)$ OPE have 
\eqn{J(x)J(0)\sim \frac{c_{JJ\CO}}{x^{4-\Delta _\CO}}\left(1+\half x^\mu \partial _\mu +\frac{\Delta _\CO +2}{8(\Delta _\CO +1)}x^\mu x^\nu \partial _\mu \partial _\nu -\frac{\Delta _{\CO}}{16(\Delta _{\CO }^2-1)}x^2\partial ^2+\cdots\right)\CO (0).}[dexp]
Explicit calculation indeed verifies, for example (in position space, using dimensional regularization), that the Wilson coefficients of the operators $\CO _n=(X^\dagger X)^n$ for the first two terms in \dexp\ indeed have the relative factor of $\half$ of \dexp; this gives a check of conformal covariance of the OPE.     

We now outline similar explicit checks of our proposed superconformal covariance of the OPE Wilson coefficients, with the generator $S$ and $\bar S$.  The proposed superconformal covariance yields many individual relations, which when combined determine the superconformal descendant Wilson coefficients in terms of those of the superconformal primaries. 

As an example, the superconformal algebra implies that 
\eqn{\bar{Q}\bar{S}(J(x)J(0))=-ix\cdot \bar{\sigma}^{\dot{\alpha}\alpha} j_\alpha(x)\bar{\jmath}_{\dot{\alpha}}(0)-2ix_\mu(j^\mu(x)-i\partial^\mu J(x))J(0).}[QbarSbar]
Taking the Fourier transform of \QbarSbar\ and using the $J(x)J(0)$ OPE \JJXterms\ and $j_\alpha (x)\bar \jmath_{\dot \alpha}(0)$ OPE \JaXterms\ yields the relation
\eqn{\tilde{d}_0(m-1,n-1;s)-\partial\tilde{d}_0(m-1,n;s)=\tfrac{1}{4}[2(1-m)-n]\left[\tilde{c}_{0}(m,n;s)-\tilde{c}_{1/2}(m,n;s)\right],}[DerRel]
where $\partial\tilde{d}_0(m,n;s)$ is the Wilson coefficient of $(F^\dagger F)^m(X^\dagger X)^n (\partial_\mu\chi)\sigma^\mu\bar{\chi}$ in \JJXterms.  Explicit computation of the Wilson coefficients verifies that \DerRel\ is indeed satisfied. 

The relation \DerRel\ determines the  $\tilde c_{1/2}$ Wilson coefficients in the $j_\alpha(x) \bar \jmath_{\dot \alpha}(0)$ OPE \JaXterms\ in terms of the Wilson coefficients $\tilde c_0$, $\tilde d_0$, and  $\partial\tilde{d}_0$ in the primary  $J(x) J(0)$ OPE \JJXterms.  This fits with the result \rcite{Fortin:2011nq} that all superconformal descendant current-current OPE coefficients are fully determined from those of the primaries.  In addition to relating the various OPEs of $J$'s descendants, $j_\alpha$, $\bar \jmath_\alpha$, and $j_\mu$, superconformal symmetry also implies relations among the terms on the RHS of the $J(x)J(0)$ OPE \JJXterms,   determining the Wilson coefficients of all superconformal descendants in terms of those of the superconformal primaries.  

As an example, consider the $\partial\tilde{d}_0(m,n;s)$  Wilson coefficient of $(F^\dagger F)^m(X^\dagger X)^n (\partial_\mu\chi)\sigma^\mu\bar{\chi}$, that entered in \DerRel.  Since $ [{\bar{S}}^{\dot{\alpha}},i\partial_\mu\chi\sigma^\mu\bar{\chi}]\neq 0$, these operators are not superconformal primary, so the coefficients $\partial\tilde{d}_0(m,n;s)$ are completely determined by the superconformal symmetry in terms of the other, superconformal primary Wilson coefficients.  Indeed, inserting the $J(x)J(0)$ OPE into superconformal relations like \SSJ\ and  
\eqn{
\left(Q^2+\frac{2i}{x^2}Q x\cdot\sigma \bar{S}\right)(J(x)J(0))=0
}
yields enough relations to, for example, fully determine the Wilson coefficients of superconformal descendants like $i\partial_\mu\chi\sigma^\mu\bar{\chi}$, $\chi x\cdot \sigma \bar{\chi}$ and $X^\dagger F^\dagger \chi^2$ in terms of the superconformal primaries. One can append $(X^\dagger X)^n (F^\dagger F)^m$ in front of all of these operators and the result remains.\foot{Since the action of $S$ on $F$ and $F^\dagger$ gives zero at $x=0$, one has to use derivative operators in order to generate the $F^\dagger F$s. Then, one can use the known action of $K_\mu$ to show that Wilson coefficients of superconformal descendants are determined in terms of those of superconformal quasi-primaries.} As expected from the analysis of \rcite{Fortin:2011nq}, the Wilson coefficients of all superconformal descendant operators are determined from those of the superconformal primaries. 

\subsec{Soft masses}

We now apply general expressions \SoftApprox\ to  analyze the gaugino and sfermion masses in this simple model.  The expressions \SoftApprox\ and \SoftApproxLow\ can be applied to strongly coupled theories, and here we verify that our techniques can indeed properly approximate soft masses in simple weakly-coupled models, where the answer is already known:  $M_{\text{gaugino}}=\frac{\alpha}{4\pi}\frac{F}{X} g(x)$ and $m_{\text{sfermion}}^2 =2 \left(\frac{\alpha}{4\pi}\right)^2 c(r) f(x)$ \rcite{Dimopoulos:1996gy, Martin:1996zb}, with  $x\equiv |F/h X^2|$ and
\eqna{g(x)&=\frac{1}{x^2}[(1+x)\ln(1+x)+(1-x)\ln(1-x)],\\
f(x)&=\frac{1+x}{x^2}\left[\ln(1+x)-2\Li_2\left(\frac{x}{1+x}\right)+\frac{1}{2}\Li_2\left(\frac{2x}{1+x}\right)\right]+(x\to-x).
}[gfare]
We find, perhaps surprisingly, that the OPE methods---generally an approximation---here reproduce the full, exact functions $g(x)$ and $f(x)$!  We discuss here the gaugino mass in some detail.  The sfermion mass computation is conceptually essentially the same, although technically a bit more involved.  

The Wilson coefficients entering in \SoftApprox\ are the $\tilde c^{\,\prime}_{1/2}(m, n; s, \mu)$ in \CoeffRel, whose imaginary parts give the discontinuity in \Bdmatch.  So \SoftApprox\ gives
\eqna{M_{\rm gaugino}\approx \alpha \sum _{m, n} \frac{\Disc [ s^{n+2m+1}\tilde c^{\,\prime}_{1/2}(m, n; s, \mu)]}{4^{n+2m+1}(n+2m+1) m_0^{2(n+2m)-1}}
 (\vev{F^\dagger F})^m(\vev{X^\dagger X})^n.}
Using the result for $\tilde c^{\,\prime}_{1/2}(m, n; s, \mu)$ in \Bdmatch, this gives $M_{\rm gaugino}\approx M_{\rm gaugino, OPE}\equiv \frac{\alpha}{4\pi}\frac {F}{X}g_{\text{OPE}}(x)$, with \eqn{
g_{\rm OPE}(x)=\sum_{n,m=0}^\infty\frac{\Gamma[2(n+m)]}{4^{n+2m}(n+2m+1)\Gamma(n)\Gamma(n+1)\Gamma(2m+2)}x^{2m}.
}
The ratio test shows that the $\sum _m$ sum converges (for $x<4$, 
which is satisfied since we anyway need $0<x<1$ to avoid tachyons), but the $\sum _n$ requires a continuation to converge.  Indeed, the $\sum _n$ sum can be rewritten  in terms of hypergeometric functions, giving
\eqn{
g_{\rm OPE}(x)=\frac{1}{2}+\sum_{m=0}^\infty\frac{1}{2^{4m+3}(m+1)}\,\phantom{}_3F_2\!\left[\genfrac{}{}{0pt}{}{m+3/2,m+1,2m+2}{2,2m+3};1\right]x^{2m}.
}[hyperi]
The hypergeometric function  $\phantom{}_3F_2\left[\genfrac{}{}{0pt}{}{a,b,c}{d,e};z\right]$ converges at $z=1$ only if $\Rea s>0$ where $s=d+e-(a+b+c)$, and that is not satisfied in \hyperi.  Fortunately, one can analytically continue the hypergeometric functions using a generalization of Dixon's theorem,
\eqn{
\phantom{}_3F_2\!\left[\genfrac{}{}{0pt}{}{a,b,c}{d,e};1\right]=\frac{\Gamma(d)\Gamma(e)\Gamma(s)}{\Gamma(a)\Gamma(b+s)\Gamma(c+s)}\,\phantom{}_3F_2\!\left[\genfrac{}{}{0pt}{}{d-a,e-a,s}{s+b,s+c};1\right],
}
which leads to convergent hypergeometric functions and gives
\eqn{
g_{\rm OPE}(x)=1+\frac{1}{6}x^2+\frac{1}{15}x^4+\frac{1}{28}x^6+\cdots=g(x).
}
The approximate $g_{\rm OPE}(x)$ function obtained from the OPE gives the exact function $g(x)$!   Similarly, the OPE approximation for the sfermion mass function $f_{\rm OPE}(x)$ actually gives the full, exact result in \gfare.  

Recalling the approximations made in \SoftApprox, it is perhaps surprising that the OPE manages to reproduce the exact results (at least in this example).  In particular, \SoftApprox\ was obtained by  approximating that there is a single cut, starting at the supersymmetric threshold for particle production, with supersymmetry breaking neglected.  We know from our discussion in  subsection \Analyticity, that this is at best an approximate oversimplification, since the 
different contributions to the soft masses actually have different cut structures.  It is interesting and curious that, at least in the present example, the OPE conspires in such a way to somehow fully account for the true cut structure.    We do not know if this occurs more generally.

Before concluding, it is interesting to see how good the approximation is if we keep only the leading order contribution \SoftApproxLow.  Using the classical OPE coefficient \ClCoeffK and the Konishi current mixing \KXXMixing, which are $1/2\pi^2$ and $|h|^2/4\pi^2$ respectively, the soft SUSY breaking functions \gfare\ can be approximated by $g(0)= f(0)\approx \half$.  
 Thus, to lowest order the approximations \SoftApproxLow allow the computation of the soft SUSY breaking parameters to an accuracy of $50\%$.  This is probably the best (and often the only) approximation to the soft SUSY breaking parameters one can achieve in strongly-coupled theories.

%%%%%%%%%%%%%%%%%%%%%%%%%%%%%%%%%%%%%%%%%%%%%%%%%%%%%%%%%%%%%%%%%%%%%%%%%%%%%%%%%%%%%%%%%%%%%%%%%%%%%%%%%%%%%%%%%%%%%%%%%%%%%%%%%%%%%%%%%%%%%%%%%%%%

\newsec{Conclusion}[CONC]

Conformal theories are interesting arenas for exploring quantum field theory.  Various possible model-building applications of approximate conformal symmetry and non-weakly coupled sectors have been proposed in the literature over the years, to help naturalize hierarchies, e.g.\ that of technicolor, flavor \rcite{Nelson:2000sn},  sequestering \rcite{Luty:2001zv}, and the $\mu$/$B\mu$ problem \rcite{Roy:2007nz, Murayama:2007ge}.  These and other models have recently motivated renewed interest in exploring the consequences of conformal or superconformal symmetry, see e.g.\ \rcite{Rattazzi:2008pe} and following papers.   Here we explore possible vestiges of approximate superconformal symmetry in wider classes of models, where the symmetries can be (softly or spontaneously) broken. 

In weakly coupled models, one can simply write down integral expressions for the GGM functions $C_a$ and $B_{1/2}$, see \rcite{Marques:2009yu, Dumitrescu:2010ha}.  Our methods here give some approximate tools to analyze theories that are not necessarily weakly coupled, giving some approximate insights on connecting the model theory to observational consequences.  It would be interesting to apply the methods to concrete examples of non-weakly coupled theories, and to explore concretely some of the above mentioned proposed applications.

%%%%%%%%%%%%%%%%%%%%%%%%%%%%%%%%%%%%%%%%%%%%%%%%%%%%%%%%%%%%%%%%%%%%%%%%%%%%%%%%%%%%%%%%%%%%%%%%%%%%%%%%%%%%%%%%%%%%%%%%%%%%%%%%%%%%%%%%%%%%%%%%%%%%
\ack{We thank Ben Grinstein, Zohar Komargodski, Aneesh Manohar, and David Shih for discussions and comments.  This research was supported in part by UCSD grant DOE-FG03-97ER40546.   KI thanks the IHES for hospitality while some of this work was done, and the participants of the IHES Three String Generations and the CERN SUSY Breaking workshops for their comments. JFF thanks the participants of SUSY11 for comments.}

\appendix

\begin{figure}[ht]
\centering
\includegraphics{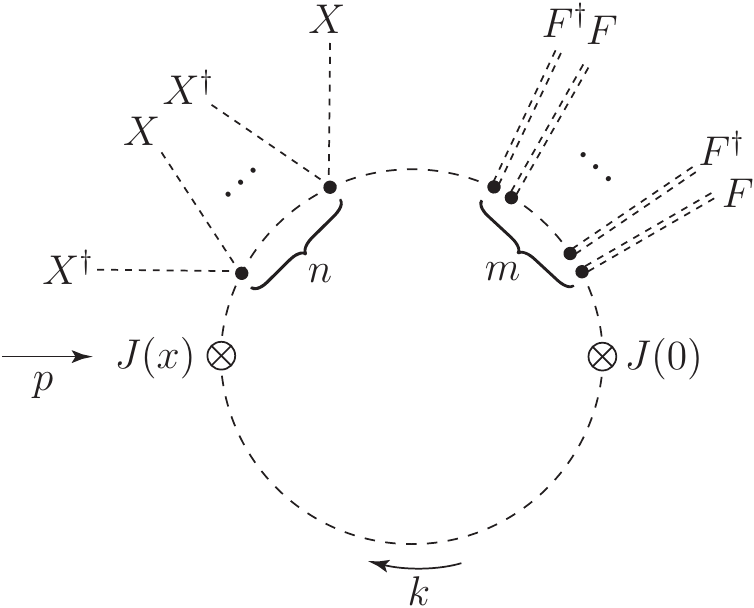}
%\pstool{./figures/diag}{\psfrag{L}[][]{$J(0)$}
%\psfrag{J}[][]{$J(x)$}
%\psfrag{p}[][]{$p$}
%\psfrag{k}[][]{$k$}
%\psfrag{m}[][]{$m$}
%\psfrag{n}[][]{$n$}
%\psfrag{X}[][]{$X$}
%\psfrag{S}[][]{$X^\dagger$}
%\psfrag{F}[][]{$F$}
%\psfrag{D}[][]{$F^\dagger$}}
\vspace{-0.2cm}
\caption{The Wilson coefficient of $(F^\dagger F)^m(X^\dagger X)^n(0)$ in the OPE $i\int d^4 x\, e^{-ip\cdot x}J(x)J(0)$.% (The Wilson coefficient is actually equal to $i$ times the corresponding amputated Feynman diagram.)
}
\label{FigDiag}
\end{figure}

\newsec{Combinatorics for Wilson coefficients}[AppWils]
In this appendix we calculate the one-loop Wilson coefficient of the operator $(F^\dagger F)^m(X^\dagger X)^n$ in the Fourier-transformed OPE of $J(x)J(0)$.  The leading contribution to the coefficient comes from the  one-loop diagram with $m$ insertions of  (the background expectation value of) $F^\dagger$ and $F$, and $n$ insertions of $X^\dagger X$ (Fig.\ \ref{FigDiag}). 
The combinatoric factors are as follows.  Permutations among the $X^\dagger X$ insertions do not count as separate diagrams, nor do permutations among $F^\dagger$s or $F$s.  $F^\dagger$ and $F$ have to be in alternating order, and only one such ordering counts.  We can start with all $X^\dagger X$ and $F$ and $F^\dagger$ insertions on the upper propagator, and then start bringing the $F^\dagger$s and $F$s, and the $X^\dagger X$s, past the current insertion, to the lower propagator. Every time an $F^\dagger$ or an $F$ goes past the current insertion, there is a minus sign, from that in 
$J(x)=\phi^\dagger \phi(x)-\tilde{\phi}^\dagger\tilde{\phi}(x)$.  After some standard manipulations for the calculation of one-loop diagrams, the Wilson coefficient computed from Fig.\ \ref{FigDiag} is  
\begin{multline*}
\tilde{c}_0(m,n;s,\mu)=\frac{|h|^{2(m+n)}}{8\pi^2}\sum_{j=0}^n\sum_{k=1+j}^{2m+1+j}(-1)^{k+j+1}\frac{(n+2m-k+1)!}{(n-j)!\,(2m-k+1+j)!}\frac{(k-1)!}{j!\,(k-1-j)!}\\
\times\int_0^1 dx\, (-1)^{n}\frac{\Gamma(2m+n)}{\Gamma(k)\Gamma(2m+n+2-k)}x^{k-1}(1-x)^{2m+n-k+1}\frac{(2m+n+1)\mu^2+\Delta}{(\mu^2+\Delta)^{2m+n+1}},
\end{multline*}
(for $m$, $n$ not both zero) where $\mu$ is the IR normalization point, and $\Delta \equiv x(1-x)Q^2$, with $Q^2=p_E^2$.  
Here $k$  counts the number of propagators that make up the lower propagator, and $j$  counts how many $X^\dagger X$ insertions are on the lower propagator.

In connection with the analyticity properties, we are particularly  interested in contribution  that is logarithmic in $Q^2/\mu^2$. Expanding the result of the above Feynman parameter integration in the UV (large $s\equiv -Q^2>0$) we get
\eqn{
\tilde{c}_0(m,n;s,\mu)\to \frac{1}{4\pi^2}\frac{(-1)^{m}\Gamma(2(m+n)-1)}{\Gamma(m+n)\Gamma(m+1)\Gamma(n+1)}\left(\frac{1}{s}\right)^m\left(\frac{|h|^2}{s}\right)^{m+n}\ln\frac{-s}{\mu^2}.
}[WilCoeff]
The case $m=n=0$, i.e.\ $\tilde c_{\mathds{1}}$, is instead given by  \momfew.  As in the discussion around \KXXMixing, the IR scale $\mu$ everywhere ultimately cancels, thanks to operator mixing, and is effectively simply replaced with the UV cutoff scale $\Lambda$.  As discussed after \Cdmatch, the combinatoric factors in \WilCoeff\ precisely reproduce the $1/s$ expansion of the kinematic factor $\lambda ^{1/2}(s, m_+, m_-)$ that enters in the cross section and the $\widetilde C_0$ discontinuity.

To outline a similar example, the Wilson coefficients $\tilde d_0(m, n; s, \mu)$ are obtained by similar considerations of a diagram like that of Fig.\ \ref{FigDiag}, but with the $X^\dagger F^\dagger \chi ^2$ external fermion insertions.  The result analogous to \WilCoeff\ is then
\eqna{\tilde{d}_0(m,n;s,\mu)\to \frac{1}{4\pi^2}\frac{\Gamma(2(m+n+1))}{\Gamma(n+1)} & \left[\frac{1}{\Gamma(2(m+2))\Gamma(n)}+(-1)^m\frac{1}{\Gamma(m+2)\Gamma(m+n+1)} \right] \\
&  \hspace{3.9cm}\times\left(\frac{1}{s}\right)^{m+1}\left(\frac{|h|^2}{s}\right)^{m+n+2}\ln\frac{-s}{\mu^2}.
}[WilCoeffd]

%%%%%%%%%%%%%%%%%%%%%%%%%%%%%%%%%%%%%%%%%%%%%%%%%%%%%%%%%%%%%%%%%%%%%%%%%%%%%%%%%%%%%%%%%%%%%%%%%%%%%%%%%%%%%%%%%%%%%%%%%%%%%%%%%%%%%%%%%%%%%%%%%%%%

\bibliography{p2ref}

\end{document}